\documentclass[preprint2]{aastex}

\usepackage[]{natbib}

\newcommand{\chandra}{{\it Chandra}}
\newcommand{\cxo}{{\it Chandra}~X-ray Observatory}
\newcommand{\etal}{{et al.}}
\renewcommand{\d}{\mbox{d}}

\received{8/28/2004}
\accepted{6/21/2005}

\shortauthors{Canizares \etal , 2005}
\shorttitle{HETG Design, Fab, Ground Cal and Flight}

\begin{document}

\title{ The \chandra\ High Energy Transmission Grating: \\
    Design, Fabrication, Ground Calibration and Five Years in Flight}

\author{Claude R. Canizares,
John E. Davis, Daniel Dewey, Kathryn A. Flanagan, Eugene B. Galton, 
David P. Huenemoerder, Kazunori Ishibashi,
Thomas H. Markert, Herman L. Marshall, Michael McGuirk, 
Mark L. Schattenburg, Norbert S. Schulz, Henry I. Smith and Michael Wise}
\email{crc@space.mit.edu} 
\affil{MIT Kavli Institute, Cambridge, MA 02139}

\begin{abstract}
Details of the design, fabrication, ground and flight calibration of the
High Energy Transmission Grating, HETG,
on the \cxo~are presented after
five years of flight experience.
Specifics include the theory of phased transmission gratings
as applied to the HETG, the Rowland design of the spectrometer,
details of the grating fabrication techniques, and the results
of ground testing and calibration of the HETG.
For nearly six years the HETG has operated essentially as designed,
although it has presented some subtle flight calibration effects.
\end{abstract}




\section{Introduction}

The \cxo\ \citep{weisskopf00},
formerly the Advanced X-ray Astrophysics Facility, AXAF,
was launched on July 23, 1999, and for five years has
been realizing its promise to
open new domains in high resolution X-ray imaging and
spectroscopy of celestial sources \citep{weisskopf04, weisskopf02}.
The High Energy Transmission Grating, HETG \citep{capella00}, is one
of two objective transmission gratings on \chandra; the
Low Energy Transmission Grating, LETG \citep{brinkman00}, is
of a similar design but optimized for energies less than 1~keV.
When the HETG is used with the
\chandra\ mirror and a focal plane imager, the resulting High Energy
Transmission Grating Spectrometer,
HETGS, provides spectral resolving powers of up to 1000
over the range 0.4-8 keV (1.5-30 \AA) for point and 
moderately extended sources.
Through year 2004, the HETGS has been used in 422 observations
covering the full range of astrophysical sources and
totaling 20~Ms, or 17~\% of Chandra observing time.
Up-to-date information on \chandra\ and the HETG
is available from the \chandra\ X-ray-Center \citep{pogv7}. 
This paper
summarizes the design, fabrication, ground and flight calibration 
of the HETG.

\begin{figure} 
\epsscale{1.0} 
\plotone{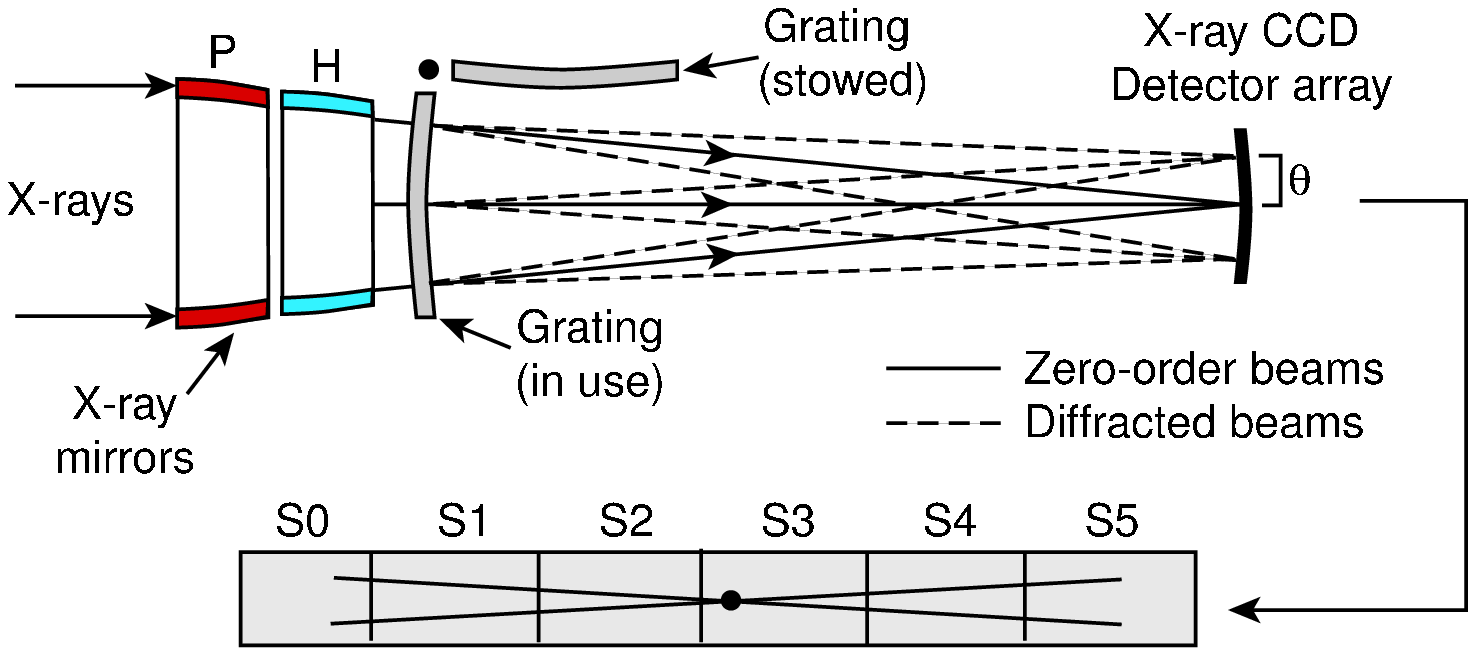}
\caption{Schematic of the HETGS on \chandra . The HETGS is formed 
by the combined operation of the mirror system (HRMA), the inserted HETG,
and the ACIS-S detector.
\label{fig:hetgs_diag}}
\end{figure}

The High Energy Transmission Grating, HETG \citep{canizares85,canizares87,
markert90,schattenburg91,markert94}, is a passive array of
336 diffraction grating facets each about 2.5~cm square.
Each facet is a periodic nano-structure consisting of finely
spaced parallel gold bars supported on a thin plastic membrane.
The facets are mounted on a precision HETG
Element Support Structure, HESS, which in turn is mounted on a hinged
yoke just behind the High Resolution Mirror Assembly,
HRMA \citep{leon97}. A telemetry command to \chandra\ activates
a motor drive that inserts HETG into the optical
path just behind the HRMA, approximately 8.6~m from the focal
plane, shown schematically in Figure~\ref{fig:hetgs_diag}. 
The lower portion of this figure shows a schematic of
the Advanced CCD Imaging Spectrometer \citep{garmire99}
spectroscopy detector, ACIS-S, 
with the HETG's shallow ``X'' dispersion 
pattern indicated.  This pattern arises from the use of two
types of grating facets in the HETG, with dispersion axes offset
by 10~$\deg$ so that the corresponding spectra are
spatially distinct on the detector.

\begin{figure}
\epsscale{1.0} 
\plotone{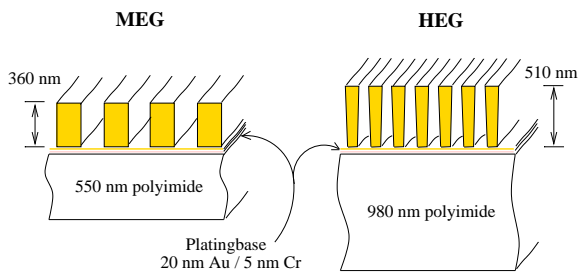}
\caption{HETG grating cross-sections.
The soap-bubble thin grating membranes of the HETG facets
consist of a supporting polyimide layer, a thin Cr/Au plating base layer,
and the actual Au grating bars.  The figures are to scale and
dimensions are approximate average values.
Note the high aspect ratio for the
HEG grating bars.
\label{fig:grat_cross_secs}}
\end{figure}

The choice (and complexity) of using two types of grating facets
resulted from our desire
to achieve optimum performance in both diffraction efficiency and spectral
resolution over the factor-of-20 energy range from 0.4 to 8~keV.
These facets are the heart of the HETG and are
schematically shown in
Figure~\ref{fig:grat_cross_secs}
with their properties given in Table~\ref{tbl-1}.
One type, the Medium Energy Grating (MEG), has
spatial period, bar thickness, and
support membrane thickness 
optimized for the lower portion of the energy range.
These MEG facets are mounted on the two outer rings of the HESS
so that they intercept rays from the 
outer two mirror shells of the HRMA, which account for $\sim$65\% of
the total HRMA effective area below 2~keV.
The second facet type, the High Energy
Grating (HEG), has finer period for higher dispersion and
thicker bars to perform better at higher energies. The HEG array
intercepts rays from the two inner HRMA shells, which have most
of the area above $\sim$5 keV.

When the HRMA's converging X-rays 
pass through the transmission grating assembly they
are diffracted in one dimension by an angle $\beta$ given
by the grating equation at normal incidence,

\begin{equation}
\sin\beta = m \lambda / p , 
\label{equ:dispersion}
\end{equation}

\noindent where $m$ is the integer order number, $\lambda$ is the
photon wavelength,
\footnote{~~
Wavelength bins are ``natural'' for a dispersive spectrometer
({\it e.g.}, the
dispersion and spectral resolution ($\Delta \lambda$) 
are nearly constant with $\lambda$) and
wavelength is commonly used in
the high-resolution X-ray spectroscopy community.  However, since photon energy
has been commonly used in high-energy X-ray astrophysics (as is 
appropriate for non-dispersive spectrometers like proportional counters
and CCDs), we use either energy or wavelength values
interchangeably depending on the context.  The values are
related through: $ E \times \lambda = 12.3985$~keV\AA.}
~~$p$ is the spatial
period of the grating lines, and $\beta$ is the dispersion angle.  A
``normal'' undispersed image is formed by the zeroth-order events,
$m=0$, while the higher orders form overlapping dispersed spectra that
stretch on either side of the zeroth-order image. By design
the first orders, $m=\pm 1$, dominate.  Higher orders are also
present however the ACIS itself has moderate energy resolution
sufficient to allow the separation of the overlapping diffracted orders.

As with other spectrometers, the overall performance of the HETGS can be
characterized by the combination of
the effective area encoded in an Auxiliary Response Function,
ARF \citep{davis01a}, and a line response
function, LRF, encoded as a Response Matrix Function, RMF.
For this grating system the LRF describes the
spatial distribution of monochromatic X-rays along the dispersion direction;
a simple measure of the LRF is the full width at half maximum (FWHM),
$\Delta\lambda_{\rm FWHM}$, expressed in dimensionless form as the
resolving power,
$\lambda /\Delta\lambda_{\rm FWHM} ~\equiv~ E/\Delta E_{\rm FWHM}$.
With the HRMA's high angular resolution (better than 1 arc sec), 
the HETG's high dispersion (as high as 100 arc
seconds/\AA), and the small, stable pixel size 
(24~$\mu$m or $\sim 0.5$~arc sec) of ACIS-S, the HETGS achieves
spectral resolving powers up to $E/\Delta E \approx 1000$.

The following sections present details of the 
key ingredients and papers related to the design of the HETGS, the fabrication
and test of the individual HETG grating facets, and the results of full-up
ground calibration of the flight HETG.
The final section demonstrates the HETGS flight performance and discusses
the calibration status of the instrument after five years of flight operation.

\section{HETG Design}

\subsection{Theory of Phased Transmission Gratings}

\begin{figure}
\epsscale{1.0} 
\plotone{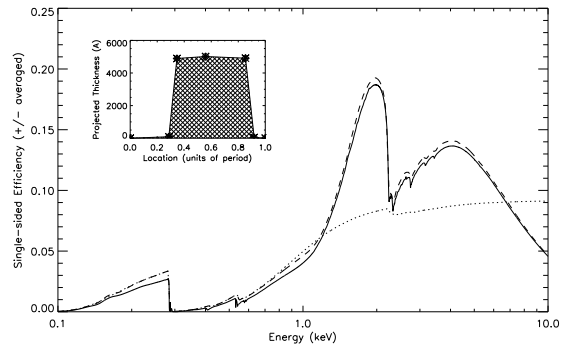}
\plotone{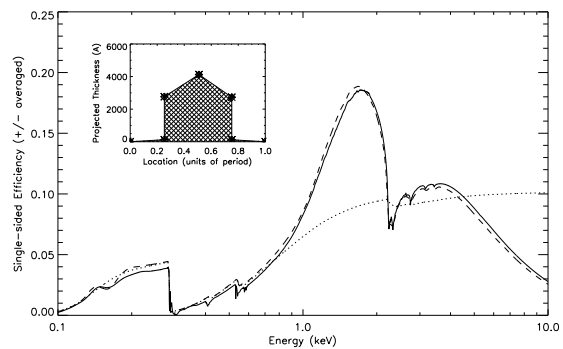}
\caption{First-order diffraction efficiencies from example
HEG (top) and MEG (bottom) multi-vertex models are plotted vs energy
(solid curves.)
The insets show the multi-vertex model grating bar cross-section.
For reference the efficiencies from a rectangular model
are shown for the cases of a constant gold thickness (dashed) and the
fully opaque case (dotted).  The enhancement of the
diffraction efficiency due to constructive phase shift which
occurs in the non-opaque cases (solid, dashed) is apparent
above 1~keV. At very high energies the non-opaque cases are introducing
less phase shift and the efficiency drops.  Note also the
subtle but significant differences between the multi-vertex efficiency (solid)
and that of a similar thickness rectangular model (dashed).
Effects of the polyimide and plating
base layers are included and produce the low-energy
fall-off and the carbon, nitrogen,
oxygen and chromium edges between 0.2 and 0.7~keV.
\label{fig:multiv_effic}}
\end{figure}

Obtaining high throughput
requires that X-rays are dispersed into the $m=\pm 1$ orders with high
diffraction
efficiency; this is largely determined by the micro-properties of the
MEG and HEG grating bars and support membranes.
Both the MEG and HEG facets are designed to operate as ``phased'' transmission
gratings to achieve enhanced diffraction efficiency over a significant
portion of the energy range for which they are optimized \citep{schnopper77}.
A conventional transmission grating with {\it opaque} grating bars achieves 
a maximum efficiency in each $\pm$1$^{st}$~order of 10\%, for the case
of equal bar and gap widths \citep[pp.~401-414]{born80}.  In contrast, the
grating bars of the MEG and HEG are 
partially transparent -- X-rays passing through the bars
are attenuated and also phase-shifted,
depending on the imaginary and real parts, respectively, of the index
of refraction at the given $\lambda$.
Ideally, the grating material and thickness can be
selected to give low attenuation and a phase-shift of 
$\approx\pi$ radians at the desired energy \citep{schattenburg84}.
This causes the radiation that passes through the bars to destructively 
interfere with the radiation that passes through the gaps when
in zeroth order (where the relevant path-lengths are equal) 
reducing the amount of undiffracted (zeroth-order) radiation and,
conversely, enhancing the diffracted-orders efficiency.
In practice, this
optimal phase interference can only be obtained over a narrow wavelength band,
as seen in Figure~\ref{fig:multiv_effic}, because of the rapid
dependence of the index of refraction on wavelength.
Initial designs based on these considerations
suggested using gold for the HEGs and silver for
the MEGs; in the end, fabrication considerations lead to
selecting gold for both grating types and 
optimizing the bar thicknesses of the MEG and HEG gratings.

Because the  grating bars are not 
opaque, the diffraction efficiency also depends on
the cross-sectional shape of the bars, 
and this shape must be determined and 
incorporated into the model of the instrument 
performance. The effect of phase shifting is
shown in Figure~\ref{fig:multiv_effic} where the dotted line
represents the single-side $1^{\rm st}$ order diffraction efficiency
({\it i.e.},
one-half the total $1^{\rm st}$ order 
diffraction efficiency) for a non-phased opaque grating
and the dashed and solid
lines are from models which include the
phase-shifting effects.
For the opaque case (dotted), the diffraction efficiency of
the grating bars is constant
with energy; the variations seen in Figure~\ref{fig:multiv_effic}
result from also including the
absorption by the support membrane and the plating base, the
thin base layers shown in
Figure~\ref{fig:grat_cross_secs}, whose
thicknesses are given in Table~\ref{tbl-1}.
These layers are nearly uniform over the grating and therefore only
absorb X-rays.

The
phased HEG and MEG gratings achieve higher
efficiencies than opaque gratings over
a significant portion of the energy band. The structure in the HEG and
MEG efficiencies is caused by structure in the index of refraction of
gold, {\it i.e.}, the M edges around 2 keV.
The efficiency falls at high energy
as the bars become transparent and introduce less phase shift to the X-rays,
falling below the opaque value at an energy depending on their thickness.
The general formula for the efficiency of a periodic transmission
grating, using Kirchhoff diffraction theory with the Fraunhofer
approximation is \citep[pp.~401-414]{born80}:
 
\begin{equation}
g_{m}(\lambda) =  {{1}\over{p^{2}}} \times \Bigl|
\int_0^p dx~e^{ik(\nu (k) - 1)z({x\over{p}}) + i 2\pi m {x\over{p}} }
 \Bigr|^{2}
\label{equ:effic}
\end{equation}
 
\noindent where $g_{m}$ is the efficiency in the $m^{\rm th}$ order,
        $k$ is the wavenumber ($2\pi /\lambda$),
        $\nu (k)$ is the complex index of refraction often
expressed in real and imaginary parts (or optical constants) as
$\nu (k) -1 = -( \delta(k) - i\beta (k))$,
        $p$ is the grating period, and
        $z(\xi) $ is the grating path-length function of the
normalized coordinate $\xi =x/p$.
The path-length function $z(\xi)$ may be thought of as the projected
thickness
of the grating bar versus location along the direction of periodicity;
at normal incidence it is simply the grating bar cross-section.

\begin{figure}
\epsscale{1.0} 
\plotone{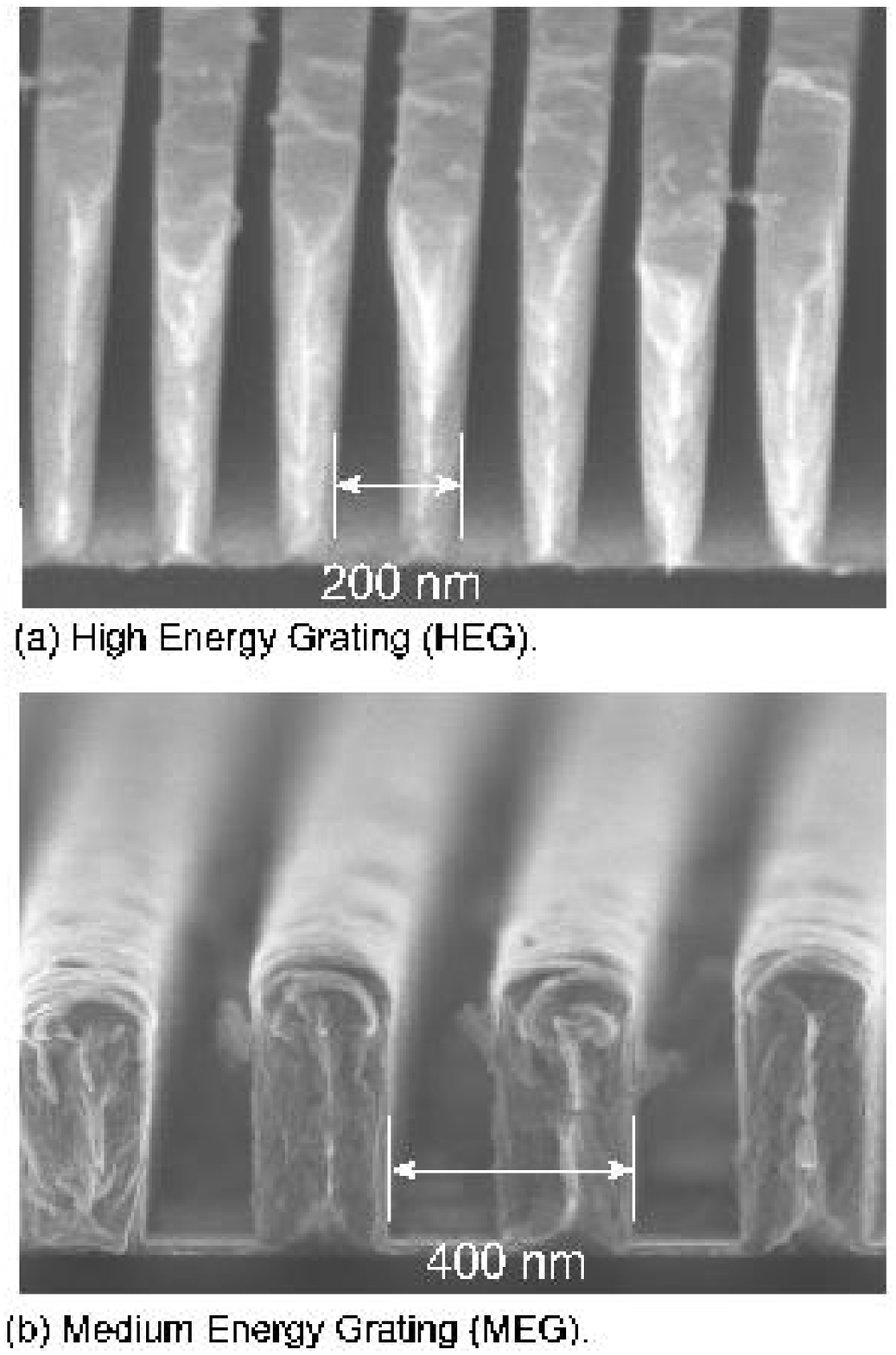}
\caption{Electron micrographs of representative HEG and MEG
grating bars.
\label{fig:emicros}}
\end{figure}

The path-length function $z(\xi)$ can be reduced to a finite number of
parameters.  For example, if a rectangular bar shape is assumed, then
$z$ can be computed with two parameters, a bar width and a bar thickness
(or height.)
Adding an additional parameter,
\citet{flanagan88} reported on the theory
and measurements of tilted rectangular gratings which yields 
a trapezoidal path-length function for small incident angles.
Our data are fit adequately for performance estimates
if a simple rectangular grating
bar shape is assumed \citep{schnopper77,nelson94}.  However,
modeling the measured first-order efficiencies to $\sim$1~\%~requires
a more detailed path-length function.  This is not
surprising given the evidence from electron microscope photographs
(Figure~\ref{fig:emicros}) that the bar shapes for the HETG gratings
are not simple rectangles.

We found from laboratory measurements (see below) 
that sufficient accuracy could be achieved by modeling
the grating bar shape
in a piece-wise linear fashion.  We parameterize the shape by specifying
a set of vertices, {\it e.g.},
as shown in the insets of Figure~\ref{fig:multiv_effic},
by their normalized locations, $\xi_j$, and
thicknesses, $z(\xi_j)$; the end-point vertices are fixed at (0,0)
and (1,0).
We have found in our modeling that 5 variable vertices are
sufficient to accurately model these gratings
in $0^{\rm th}$, $1^{\rm st}$ and $2^{\rm nd}$ orders
and yet not introduce redundant parameters.
Finally, this multi-vertex path-length
function also lends itself to simple calculation as presented in
Appendix~\ref{sec:effic_append}.

Note that this multi-vertex model allows an asymmetric
(effective) bar shape
which generally leads to
unequal plus and minus diffraction orders, as in
the case of a blazed transmission grating \citep{michette93}
or as arises
when a trapezoidal grating is used at off-normal (``tilted'') incidence.
Hence, this multi-vertex model
can be a useful extension of the symmetric, multi-step scheme
formulation in \citet{hettrick04}.
For the
HETG gratings this asymmetric
case arises primarily when the roughly trapezoidal
HEG gratings were tested at non-normal incidence, producing
an asymmetry of up to 30~\% per degree of tilt.  However, 
the asymmetry is linear for small tilt
angles from the normal and so the sum of the plus and minus order efficiencies
remains nearly constant.

\subsection{Synchrotron Measurements and Optical Constants}

We used high intensity X-ray beams at several synchrotron radiation
facilities for several purposes: (i) to measure the optical constants
and absorption edge structure of the grating materials and supporting
polyimide membranes; (ii) to make absolute efficiency measurements of several
gratings to validate and constrain our grating performance model and provide
estimates of its intrinsic uncertainties; (iii) to calibrate several 
gratings for use as transfer
standards in our in-house calibration; and (iv) to measure the efficiencies
of several gratings also calibrated in-house to assess uncertainties;
see ~\citet{flanagan96,flanagan00}.  
Synchrotron radiation tests were performed at four facilities 
over several years.

\begin{figure}
\epsscale{1.0} 
\plotone{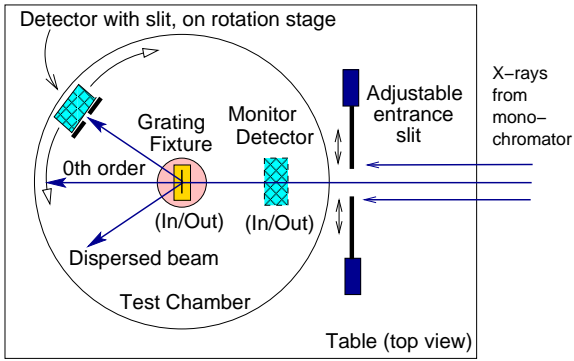}
\caption{
Measurement configuration at the National Synchrotron Light Source.
X-rays from the beam line monochromator are incident from the right
and collimated by an entrance slit.  A monitor detector can be quickly 
inserted to provide accurate normalization of the beam.  The main detector
measures the grating-dispersed X-ray flux and can be scanned in angle.
Adapted from \citet{nelson94}.
\label{fig:synchro_setup} }
\end{figure}

A first set of measurements were made at the National Synchrotron Light
Source (NSLS) at Brookhaven National Laboratory (BNL)
piggy-backing on equipment and expertise \citet{graessle96} developed
to support the determination of the reflectivity properties of the HRMA coating.
The general configuration of these tests is indicated in 
Figure~\ref{fig:synchro_setup}; key ingredients are: an input of bright,
monochromatic X-rays from the beam line, a beam monitoring
detector which is inserted frequently to normalize the beam intensity,
a detector fitted with a narrow slit (0.002 and 0.008 inches were used)
that could be rotated to intercept radiation at a desired diffraction angle,
and the grating itself which could be rotated (``tilted'')
about the vertical (grating bar) axis and also removed for detector cross-calibration.
Data sets were taken automatically with one or more of the controlled
parameters varied: monochromator energy scan, diffraction angle (order) scan, and
a grating tilt scan.

Initial modeling based on a rectangular grating bar 
model and using the optical constants, $\delta$ and
$\beta$, obtained from the scattering factors ($f_1$, $f_2$) published by
\citet{henke}  indicated
significant disagreement with the Henke values for the gold optical
constants \citep{nelson94, markert95}.
The most noticeable feature was that the energies of the
gold M absorption edges were shifted from the tabulated amounts by as
much as 40~eV, a result obtained earlier by \citet{blake}
from reflection studies of gold mirrors.  In an
effort to determine more accurate optical constants, the transmission
of a gold foil was measured over the range 2.03--6.04~keV, and the
values of $\beta$ and $\delta$ were revised \citep{nelson94}.  The
widely used Henke tables were modified in 1996 to reflect these results.

Subsequent tests on gratings explored bar shape, tilt and
asymmetry \citep{markert95}, and tests at the radiometry laboratory of
the Physikalish-Technische Bundesanstalt (PTB) below 2~keV identified
the need to accurately model the edge structures of the polyimide
support membrane to improve the overall fit \citep{flanagan96}. The
analysis of the tests of gold and polyimide membranes at PTB
in October 1995 is detailed in \citet{flanagan00}.
In addition, cross-checks of the
revised gold constants (above 2~keV) and polyimide were performed in
August and November 1996 and have confirmed these latest revisions.

\begin{figure}
\epsscale{1.0} 
\plotone{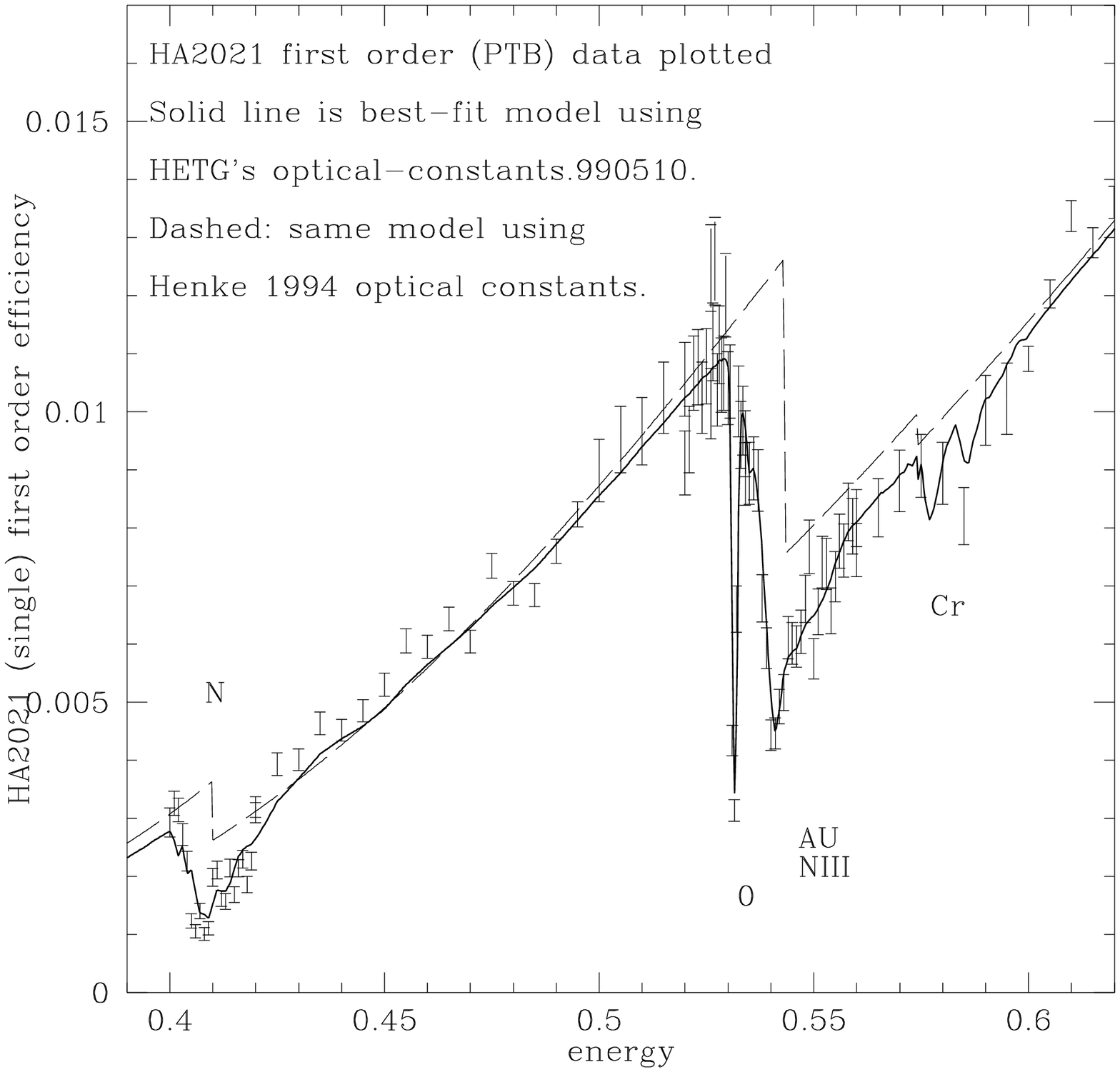} \\
\plotone{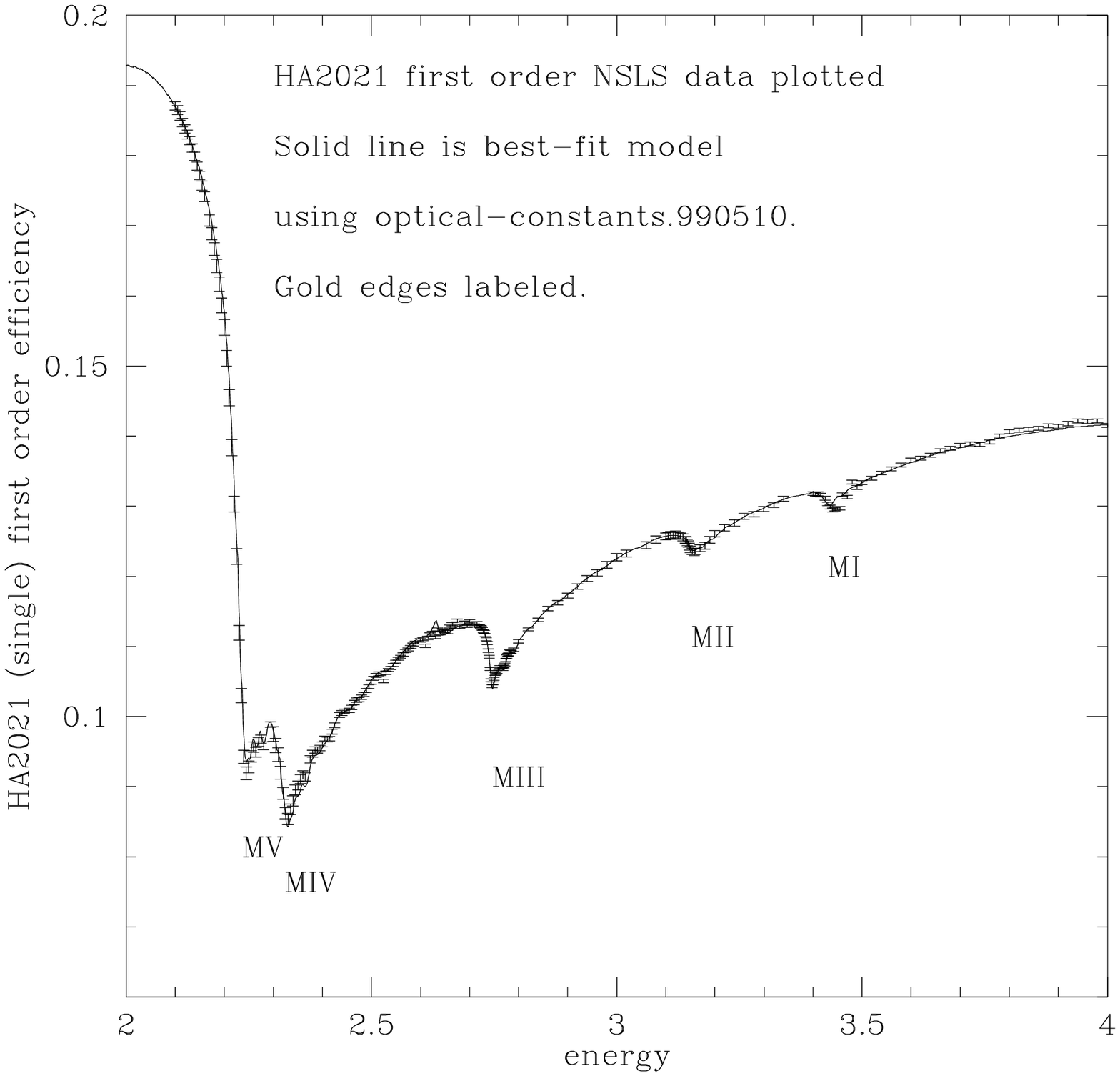} \\
\caption{Synchrotron efficiency measurements.
The first-order efficiency is generally smoothly varying with
energy in the HETGS band (see previous figure)
except in the 
polyimide and Cr edge region (top) and the
gold M edge region (bottom).
The modeling process was driven by
extensive sample measurements made with monochromatized
synchrotron light sources.  Shown here are the
finely spaced measured data (error bars)
with a best-fit multi-vertex model (solid line).
\label{fig:synch_effic}}
\end{figure}

As a consequence of these analyses, our model now includes 
revised gold optical constants over the 
full energy range appropriate to the HETG, and 
detailed structure for absorption edges 
of polyimide, C$_{22}$H$_{10}$O$_4$N$_2$, and Cr.
An example of the agreement between measured
and modeled efficiencies is shown in Figure~\ref{fig:synch_effic}.

\subsection{The Faceted Rowland Design}

The HETGS optical design is based on an
extension of the simple Rowland spectrometer design in which the
gratings and detector are located on opposite sides of an imaginary
 Rowland circle \citep{born80}. The Rowland
configuration maintains the telescope focal properties in the dispersion
direction for a large range of diffraction angle, $\beta$, thereby
minimizing aberrations.
A detailed discussion of the physics of Rowland
spectrometers, {\it i.e.}, applying Fermat's
principle \citep{schroeder87} to evaluate aberrations of the faceted
grating design, is given by \citet{beuermann78}.
What follows is a simplified, ray-based description of the basic design.

\begin{figure}
\epsscale{1.0} 
\plotone{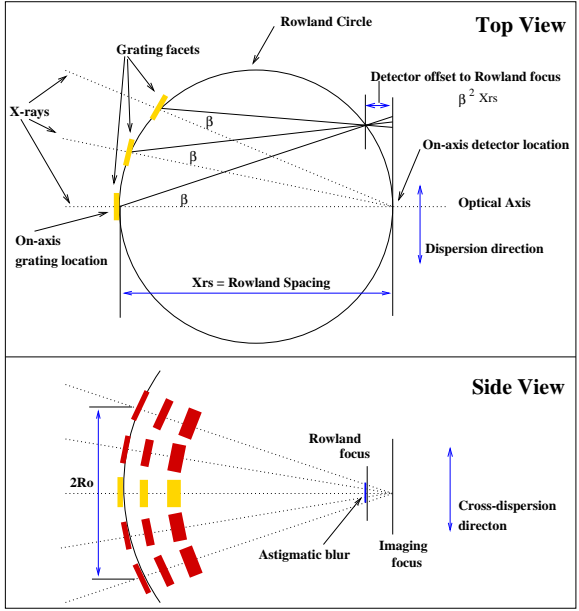}
\caption{Simplified ray geometry
for the Rowland torus design.  The {\it Top View}
shows the spectrometer layout viewed from +Z (``above \chandra '')
with the HRMA off the page to the left. X-rays
reflected by the HRMA come to a focus at zero order (dotted lines).
The grating facets diffract the rays into the m$^{\rm th}$-order
spectra at angle $\beta$ with respect to the optical axis, and
bring the dispersed spectrum to a focus on the Rowland circle (solid lines).
The Rowland spacing, $X_{\rm rs}$, is the
diameter of the Rowland circle and the distance from the gratings to
the detector.  In the {\it Side View}, we see the cross-dispersion
projection of the same rays.
Notice that in the cross-dispersion-direction, the diffracted rays focus
behind the Rowland circle.
\label{fig:rowland_design}}
\end{figure}

The ``Top View'' of Figure~\ref{fig:rowland_design} shows the plane of
dispersion (the $(x',y')$ plane), as viewed along
the cross-dispersion direction, $z'$.  The diffraction angle is
$\beta$, as defined by Equation~\ref{equ:dispersion}; note that
the facet surfaces are normal to the incoming, central X-rays
and are thus not tangent to the Rowland circle.
Through the
geometric properties of the circle, rays diffracted from gratings
located along the Rowland circle will all converge at the same
diffracted point on the Rowland circle.  The dotted lines represent
zeroth-order ($m=0,\beta = 0$) rays and the solid lines a set of
diffracted order ($m > 0, \beta > 0$) rays.

The bottom panel, ``Side View'', gives a view along the dispersion direction,
$y'$, at rays from a set of grating facets located in the the
$(x',y')$ plane (the same three facets in the ``Top View'' now seen
in projection,
shown in light shading) as well as additional
grating facets (darker facets) 
located above (or equivalently below) the $(x',y')$
plane. Each arc of additional facets is located on another Rowland circle
obtained by rotating the circle in the Top View about the
right-most line segment:
the tangent to the Rowland circle 
that is parallel to the dispersion direction and passes through
the zeroth order focal point.  The surface described by this rotation 
is the {\it
Rowland torus}.  All grating facets with centers located on this Rowland
torus and with surfaces normal to the converging rays (dotted
lines) will focus their diffracted orders on a common arc on 
the Rowland circle in the $(x',y')$ plane.  Since the
Rowland diameter is the same for all grating facets, and the zero order
focus coincides for all facets, the m$_{\rm th}$ diffracted order from
each facet is focused at the same angle $\beta$, at the same place on
the Rowland circle.  That is, best focus for the dispersion direction
projection occurs along the inner surface of the Rowland torus.

\begin{figure}
\epsscale{1.0} 
\plotone{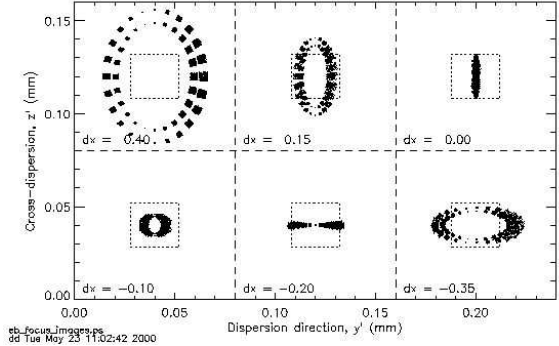}
\caption{Ray-trace of Faceted Rowland Geometry:
19~\AA~MEG images vs defocus.
The focal properties of the faceted Rowland design are
demonstrated in this set of images at different defocus
values, $dx$; positive values are a displacement of the detector
towards the grating.  Parameters of the simulation approximate
the MEG gratings on \chandra\ at a wavelength of 19~\AA. 
At large defocus values ($dx$=0.40~mm) the rays from each facet
are visible, here there are 24 facets in each of two shells.
The image comes to a minimum width in the dispersion direction
at the Rowland focus, $dx=0$, with a finite cross-dispersion
width.
At a defocus of $dx\approx -0.20$ the local detector surface
is in the focal plane and the image is now well-focused
in the cross-dispersion direction. 
These spot diagrams were created by simple ray-tracing
of a perfect focusing optic combined with a faceted
grating set -- hence, the inherent astigmatism and finite facet-size
blur of the Rowland design dominate the image at best focus,
$dx = 0.0$.
For reference, the dotted square is the size of an ACIS pixel.
\label{fig:rowland_focus}}
\end{figure}

Together, these constructions show the astigmatic nature of the
dispersed image: the rays come to a focus in the dispersion
direction, the {\it Rowland focus}, at a different location from their
focus in the cross-dispersion direction, the {\it Imaging focus}.
This is demonstrated in the ray-trace example of
Figure~\ref{fig:rowland_focus}.

In order to maintain the best spectroscopic focus the detector surface
must conform to or approximate this Rowland curvature so that
diffracted images are focused and sharp in the dispersion direction,
and elongated in the cross-dispersion direction.  The offset of 
the Rowland circle from the tangent at the zeroth order focus is
\begin{equation}
  \Delta X_{\rm Rowland} = \beta^2 X_{\rm RS}
\label{equ:delta_x_rowland}
\end{equation}
\noindent where $X_{\rm RS}$ is the Rowland spacing, the diameter
of the Rowland circle.

At the Rowland focus, {\it e.g.}, the $dx=0$ case in
Figure~\ref{fig:rowland_focus}, the image is elongated (blurred) in the
cross-dispersion direction, $z'$, due to the astigmatic nature
of the focus and has a peak-to-peak value given by:
\begin{equation}
  \Delta z'_{\rm astig} = {\frac{2R_0}{X_{\rm RS}}}   \Delta X_{\rm Rowland}
\label{equ:delta_zprime}
\end{equation}
\noindent Here $R_0$ is the radius of the ring of
gratings around the optical axis as defined in
Figure~\ref{fig:rowland_design}.  The width of the image in the
dispersion direction, $y'$, is given by a term proportional to the
size of the (planar) grating facets which tile the Rowland torus.
The peak-to-peak value of this ``finite-facet size'' blur is
given by:
\begin{equation}
\Delta y'_{\rm ff} = \frac{L}{X_{\rm RS}}  ( R_0\beta
			+  \frac{\Delta X_{\rm Rowland}}{2} )
\label{equ:ff_ppblur}
\end{equation}
\noindent where $L$ is the length of a side of the square grating facet.
This blur sets the fundamental resolving power limit
for the Rowland design with finite-sized facets.
For the HETGS design this contribution
is much smaller than the ACIS pixel size,
see the $dx=0$ case of Figure~\ref{fig:rowland_focus},
and is negligible compared to the terms in the 
resolving power error budget of 
Appendix~\ref{sec:rowland_append}.

\section{HETG Fabrication, Test, and Assembly}

\subsection{Facet Fabrication}

The 144 HEG and 192 MEG grating facets are the key components of the HETG
and presented major technical challenges: to create facets with 
nanometer scale periods, nearly rectangular bar shape, 
nearly equal bar and gap widths, and
 sufficient bar depth to achieve high diffraction efficiencies,
and with a high degree of uniformity in all these properties within each
facet and among
hundreds of facets. The facets must also be sufficiently robust
to withstand the vibrational and acoustic rigors of space launch without
altering their properties, much less being destroyed.

The HETG grating facets were fabricated, one at a time, 
in an elaborate, multi-step
process employing techniques adapted from those used to fabricate
large-scale integrated circuits.  Development work was
initiated in the Nano Structures Laboratory and
refinement of these processes
took place over nearly two decades with final flight
production in the Space Nanotechnology Laboratory
of the (then) MIT Center for Space Research.

\begin{figure}
\epsscale{1.0} 
\plotone{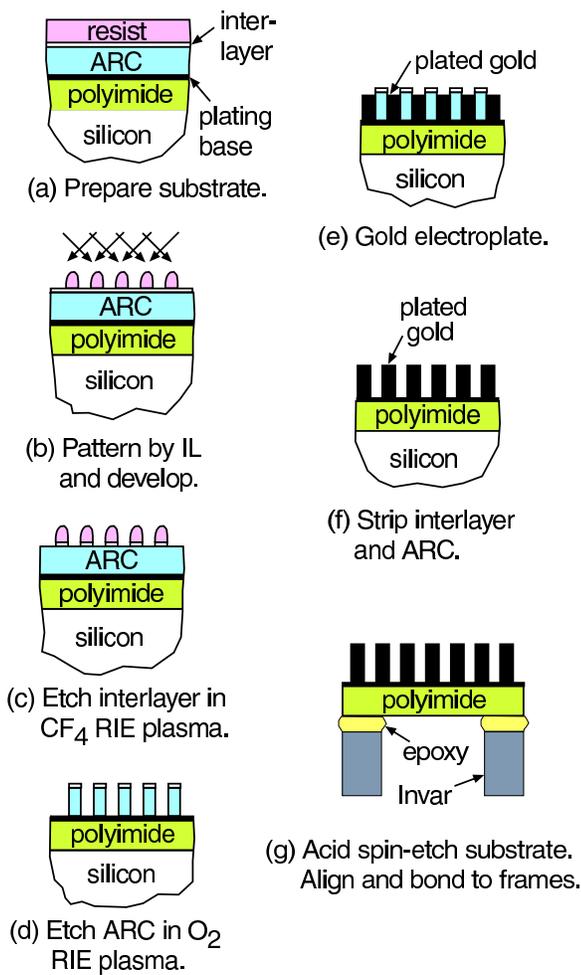}
\caption{Simplified
production steps for the HETG facets.
The initial periodic pattern is created as the interference of two
laser wave fronts.  This pattern is etched into the polymer.  Through
electroplating gold is deposited into the spaces between polymer bars.
Removal of the polymer (stripping) and Si wafer support leaves the grating
membrane in the wafer center.  This membrane is then aligned and bonded to
the Invar frame.
\label{fig:fab_steps}}
\end{figure}

Each facet was fabricated on a silicon wafer, which was used as a substrate
but did not form any part of the final facet.  In brief, the process
involved depositing the appropriate material layers on the wafer,
imprinting the period on the outermost layer using UV laser interference,
transferring that periodic pattern to the necessary depth, thereby creating
a mold with the complement of the desired grating geometry, filling
the mold with gold using electroplating, stripping away the mold material,
etching away a portion of the Si substrate, aligning and attaching a frame and
then separating the finished facet from the silicon wafer. A highly simplified
depiction of these steps is given in Fig.~\ref{fig:fab_steps} and described
below.  More complete details of the fabrication process
are available elsewhere \citep{schattenburg94,schattenburg01}.

The first step, Fig.~\ref{fig:fab_steps}a, is to coat 100~mm-diameter
silicon wafers with six layers of polymer, metal, and dielectric,
comprising either 0.5~(MEG) or 1.0~(HEG) microns of polyimide (which
will later form the grating support membrane), 5 nm of chromium (for
adhesion) and 20~nm of gold which serve as the plating base,
$\approx$500~nm of anti-reflection coating~(ARC) polymer, 15~nm of
${\rm Ta}_2{\rm O}_5$ interlayer~(IL), and 200~nm of UV imaging
photopolymer (resist).

The second step, Fig.~\ref{fig:fab_steps}b, is to expose the resist
layer with the desired periodic pattern of the final grating 
using interference lithography at a wavelength of 351.1~nm. 
Two nearly spherical, monochromatic wavefronts
interfere to define the grating pattern
period; the radii of the spherical wavefronts 
are sufficiently large to reduce the inherent period
variation across the sample to less than 50~ppm~rms.  A high degree of
period repeatability is required from the hardware because a unique
exposure is used for each grating facet of the HETG. Prior to each
exposure, the Moire\'e pattern between the UV interference standing waves
 and a stable reference
grating fabricated on silicon was used to lock the  
interferometer period.   A secondary
interferometer and active control are used to ensure that the
interference pattern is stable over the approximately one minute exposure
time.  The interlayer, ARC, and resist layers form an
optically-matched stack designed to minimize the formation of planar
standing waves normal to the surface, which would compromise
contrast and linewidth control \citep{fabrefE}.

In the third step, Fig.~\ref{fig:fab_steps}c, the resist pattern is
transferred into the interlayer using ${\rm CF}_4$ reactive-ion plasma
etching (RIE).  In the fourth step, Fig.~\ref{fig:fab_steps}d, the IL
pattern is transferred into the ARC using ${\rm O}_2$ RIE.  The RIE
steps are designed to achieve highly directional vertical etching with
minimal undercut.

The fifth step, Fig.~\ref{fig:fab_steps}e, is to electroplate the ARC
mold with low-stress gold, which builds up from the Cr/Au plating base
layer.  The sixth step, Fig.~\ref{fig:fab_steps}f, is to strip the
ARC/IL plating mould using hydrofluoric (HF) acid etch, and plasma
etching with ${\rm CF}_4$ and ${\rm O}_2$.  At this point the gold
grating bars are complete; Figure~\ref{fig:emicros}
shows electron micrographs of cleaved cross-sections of the gratings.

\begin{figure}
\epsscale{1.0} 
\plotone{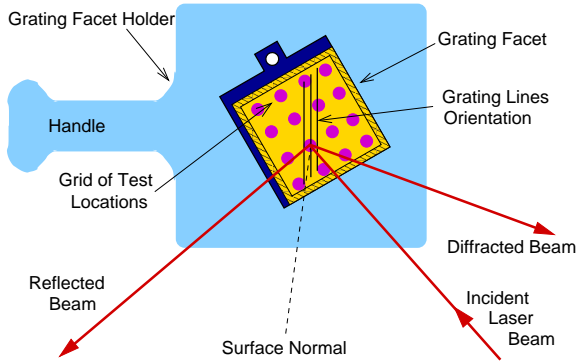}
\caption{
Laser Reflection, LR, principle of measurement.
Note that during this and other laboratory testing, the flight
grating is not directly handled: the {\it non-flight} grating facet
holder serves as interface to both humans and equipment.
\label{fig:lr_setup} }
\end{figure}

In the last step, Fig.~\ref{fig:fab_steps}g, a circular portion of the 
Si wafer under the grating and membrane is etched
through from the backside in ${\rm HF/HNO}_3$ acid using a spin etch
process that keeps the acid from attacking the materials on the front side
 \citep{fabrefF}.  The
membrane, supported by the remaining ring of un-etched Si wafer,
is then aligned to an angular tolerance of
$\leq 0.5$~degree and bonded to a flight ``frame'' using a two-part,
low-outgassing epoxy.  Once cured, the excess membrane and Si ring is
cut away from the frame with a scalpel. 
The frames are custom-made of black chrome plated Invar 36,
machined to tight
tolerances, and the membrane bonding faces were hand
lapped to remove burrs and ensure a flat, smooth surface during
bonding.
Use of Invar reduces any grating period variations which might be caused by
thermal variation of the HETG environment between stowed and in-use
positions on \chandra .  
Likewise, the frame design has a single mounting hole to reduce the effect of
mounting stresses on the facet period.
Each completed
facet was mounted in a non-flight ``holder'' to allow for ease in storing,
handling, and testing; a schematic of holder and facet is shown
in Figure~\ref{fig:lr_setup}.

After years of preparation, we fabricated 245 HEG and 265 MEG
gratings in 21 lots over a period of 16 months; tests on the
individual facets, next section, were used to select the elite
set of 336 flight grating facets.
As a postscript to the fabrication of the \chandra\ HETG facets, we
note that we have since extended our 
technology \citep{schattenburg01} to fabricate
gratings with finer periods \citep{fabrefB}, 
mesh-supported ``free-standing''
gratings for UV/EUV and atom beam diffraction and filtering \citep{fabrefC},
and super-smooth reflection gratings \citep{fabrefD}.

\subsection{Facet Laboratory Tests \& Calibration}

The completed HETG facets were put through a set of
laboratory tests to characterize their quality and performance and to enable
selection of an optimal complement of flight gratings. 
Each facet went through a sequence of tests: i) Visual Inspection, ii)
Laser Reflection (LR) Test 1,
iii) Acoustic Exposure, iv) LR Test 2, v) Thermal Cycling,
vi) LR Test 3, and finally vii) X-ray Testing.
As noted, to reduce direct handling each fabricated facet was mounted to its own
aluminum holder, the facet-level test equipment was designed
to interface to the holder.

The laser reflection, LR, test \citep{dewey94} uses optical diffraction of a 
laser beam (HeNe 633nm for MEG, HeCd 325 nm for HEG) from the grating surface
to measure period and period variations of each facet.
As shown in Figure~\ref{fig:lr_setup} the laser beam is incident on the
grating under test at an off-normal angle.  A specularly reflected beam
and a first-order diffracted beam emerge from the illuminated region
of the grating.  These beams are focused with simple,
long-focal length ($\approx$500~mm)
lenses onto commercial CCDs.  Under computer control the grating is moved so that
a raster of over 100 regions is illuminated and the centroids of the reflected and
diffracted beams in the CCD imagers are measured and recorded.  Changes
in the four CCD spot coordinates,
$ X_{\rm Refl}, Y_{\rm Refl}, X_{\rm Diff},
Y_{\rm Diff}$, are linearly related to changes in four local grating properties:
the grating surface tilt and tip, the grating period and
the grating line orientation (roll.)
These measurements are referenced to gratings (HEG and MEG)
on silicon substrates permanently mounted
in the system and measured before and after each raster scan set.
The LR data files are used to determine for each grating facet 
a mean period $p$ and an rms period variation $dp/p$ as well as
contours of period variation across the facet.
The flight grating sets were then selected to achieve minimal
overall period variation for the complete HEG and MEG arrays,
106 and 127~ppm~rms, respectively.  The ability
of the LR apparatus to measure absolute period 
was calibrated using samples on silicon measured independently at the
National Institute for Standards and Technology. The average
periods of the grating sets as determined from the LR measurements are given in
Table~\ref{tbl-1}, Laboratory Parameters.

\begin{figure}
\epsscale{1.0} 
\plotone{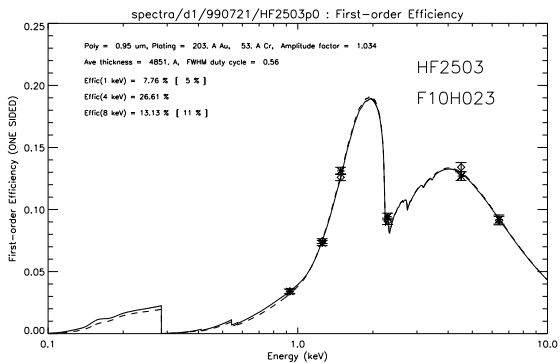}
\caption{Example of X-GEF measurements and model fit.
The measured plus and minus first-order efficiencies at the six test energies
are shown by the symbols with error bars; also plotted
are the first-order efficiencies of the best-fit multi-vertex
model, solid and dashed curves.
Second and zeroth-order measured efficiencies (not shown) have also been
included in the model fit.
\label{fig:xgef_example}}
\end{figure}

The diffraction efficiency \citep{dewey94} of each facet was measured
using the X-Ray Grating Evaluation Facility, X-GEF, 
consisting of a laboratory electron-impact
X-ray source, a collimating slit and grating
assembly, and two detectors (a position sensitive proportional
counter and a solid state detector) in a 17~m long vacuum system.
Facet tests were conducted  at a
rate of 2 gratings per day.  The zeroth, plus and minus first and second
order efficiencies were measured for five
swath-like regions on each facet
and at up to six energies, Cu-L 0.930 keV, Mg-K
1.254 keV, Al-K 1.486 keV, Mo-L 2.293 keV, Ti-K 4.511 keV, and Fe-K
6.400 keV.
Two facets, which had been
tested at synchrotron facilities \citep{markert95}, 
served as absolute efficiency
references.  
The measured monochromatic efficiencies were 
fit with our multi-vertex efficiency model \citep{flanagan95},
an example model fit to X-GEF measured points is shown in Figure~\ref{fig:xgef_example}.
These measurements and models allowed us to select the highest efficiency
gratings for the flight HEG and MEG sets as well as to predict
the overall grating-set efficiencies, Section~\ref{sec:efficEA}.

The HETG flight-candidate gratings were X-GEF tested from mid-1995
through September of 1996 at a typical rate of two per day. 
A small set of non-flight gratings were retained in a laboratory
vacuum and their diffraction efficiencies were measured with
X-GEF at
seven epochs from late 1996 to February 2003.  These ``vacuum storage
gratings'' showed no evolution in their diffraction properties
giving us an expectation of stability for the HETG efficiency
calibration.

\subsection{HETG Assembly}

\begin{figure} 
\epsscale{1.0} 
\plotone{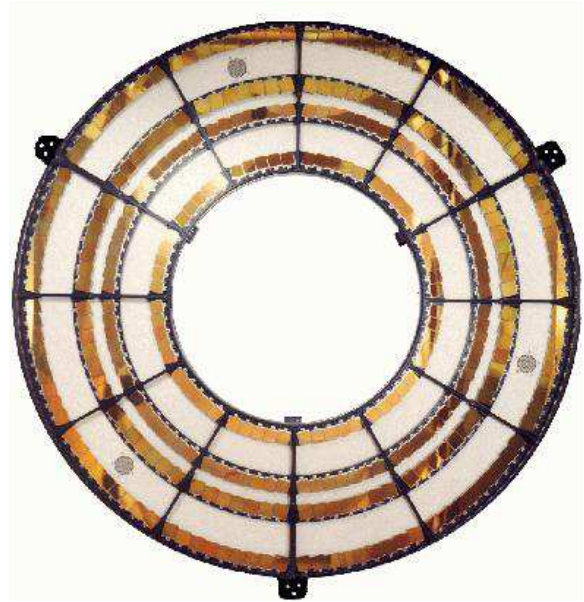}
\caption{Photograph of the HETG. Glittering gold, the 336
grating facets are visible mounted to the black-anodized support
structure, the HESS.  The outer two rings of gratings are MEGs
and intercept rays from the HRMA shells ``1'' (outer most) and ``3'';
the inner two rings are HEGs and work with HRMA shells ``4''
and ``6'' (inner most.) 
\label{fig:hetg_photo}}
\end{figure}

The flight HETG came into being when
the selected 336 flight grating facets were mounted to the
HETG Element Support Structure, HESS.
The HESS was
numerically machined from a single plate of aluminum $\approx 4$~cm thick
 \citep{pakmcguirk94,markert94}
to create a spoke-and-ring structure with mounting surfaces and holes
for the facets that
conformed to the Rowland torus design with a diameter
given in Table~\ref{tbl-1}.
The HESS mechanical design using tapered $\approx 6$~mm thick
spokes achieves the objectives of: a low
weight, an accurate positioning of the facets, and 
the ability to withstand the high-$g$ launch vibration environment.
The flight HETG is shown in Figure~\ref{fig:hetg_photo} where the
HESS is black and the facet surfaces are gold;
its outer diameter is 1.1~m and
three attachment points provide for its mounting to
one of the two \chandra\ telescope grating insertion mechanism yokes.
The completed flight HETG weighs 10.41~kg of which 8.88~kg is due to the
HESS structure, 1.21~kg for the grating elements, and 0.32~kg for the
the element-to-HESS mounting hardware.  The ``active
ingredient'' of the HETG, the gold grating bars, weighs a meager 1.14~mg.

The single-screw mounting scheme
used to attached the facets to the
HESS adequately
fixes all degrees of freedom of the facet except for rotation
around the screw axis, {\it i.e.}, the ``roll'' angle of the facet.
The roll angle was aligned using the ability of the grating to polarize
transmitted light which has a
wavelength longer than the grating period.
A schematic of our setup, based on the polarization alignment 
technique of \citet{anderson88}, is shown in Figure~\ref{fig:align_setup}.
Light from the HeNe laser passes through a photo-elastic modulator
at a 45~degree angle.  The emerging beam can be viewed as having two linearly
polarized components at right angles with a time dependent relative
phase varying as $sin(\omega t)$.
Ignoring any effect of the polyimide on the light, the polarizing grating
bars transmit only the projection of these components that is perpendicular
to the bars.
For a non-zero $\theta_g$ some
fraction of each of the modulator-axes components is transmitted resulting
in interference and an intensity signal at $2\omega$ proportional to $\theta_g$.
This measurement setup was used along with appropriate manipulation
fixturing to set each facet to its
desired roll orientation (differing by $\approx 10$~degrees
between HEG and MEG facets) with, in general, an accuracy better than 1~arc~minute. 

When all facets were aligned and the alignment re-checked
they were then epoxied to the HESS.
The flight HETG was then subjected to a random vibration test. 
Once again, 
the alignment apparatus was used to make a set of measurement of the 
facet roll angles. These final measurements indicated that 
all gratings were held secure and the
roll variation was 0.42~arc~minutes~rms
averaged over all gratings, with less than a dozen
facets having angular offsets in the  1--2.2~arc~minute range. 
During subsequent full-up ground calibration
using X-rays, next section, we discovered that, in fact, the roll angles of
6 of the MEG facets showed improper alignment.

\section{Pre-Flight Performance Tests \& Calibration}

The \cxo\ components most relevant to flight performance were tested at
the NASA Marshall Space Flight Center X-Ray Calibration Facility, XRCF, in
Huntsville, AL from late 1996 through Spring of 1997
\citep{weisskopf97,odell98}.  These full-up tests provided unique information
on the HETG and its operation with the HRMA and ACIS.  Key results of
this testing are summarized here; details of the analyses are in
the cited references and the HETG Ground Calibration Report \citep{CalReport}.

With the test X-ray sources \citep{jeffk95} located at a finite distance
from the HRMA, 518 meters,
the HRMA focal length at XRCF was longer by $\approx$200~mm
than the expected flight value.
In order to optimally intercept the rays exiting the HRMA hyperboloid at XRCF,
the HETG Rowland spacing, that is the distance from the HRMA focus
to the HETG effective on-axis location, was increased by $\approx$150~mm
over its design value, Table~\ref{tbl-1}.
We evaluated the effect of this difference between the as-machined Rowland
diameter, $X_{\rm HESS}$, and the as-operated Rowland spacing, $X_{RS,XRCF}$,
using our ray-trace code; this let us set the scale factor for a simple analytic estimate
of the rms dispersion blur:

\begin{equation}
\sigma_z ~ \approx ~ 0.2 ~ R^2_0 ~ {\frac{\lambda}{p}} ~ ( {\frac{1}{X_{\rm HESS}}} - 
  {\frac{1}{X_{RS,XRCF}}} ),
\label{equ:offRowland}
\end{equation}

\noindent Using extreme-case
values ($R_0=$500.0~mm, $\lambda =$40~\AA, $p=$4000~\AA)
this equation gives 
an additional blur of order one micron rms, insignificant
compared to the image rms which is greater than 15~$\mu$m~rms.

In addition to the HETG spacing difference, 
other aspects of the XRCF testing differed from flight 
conditions.
The HRMA, which was designed to operate in a 0-g
environment, was specially supported and counter-balanced to operate 
in 1-g.  This results
in a mirror PSF that is not identical to the PSF expected in flight.
A non-flight shutter assembly allowed quadrants
of the HRMA shells to be vignetted as desired; among other things,
 this allowed
the HEG and MEG zeroth orders to be  measured independently.
In addition to the flight detectors,
ACIS and High Resolution Camera, HRC \citep{murray98},
several specialized detectors were used to conduct
the tests.

\subsection{Line Response Function Measurements}

\begin{figure}
\epsscale{1.0} 
\plotone{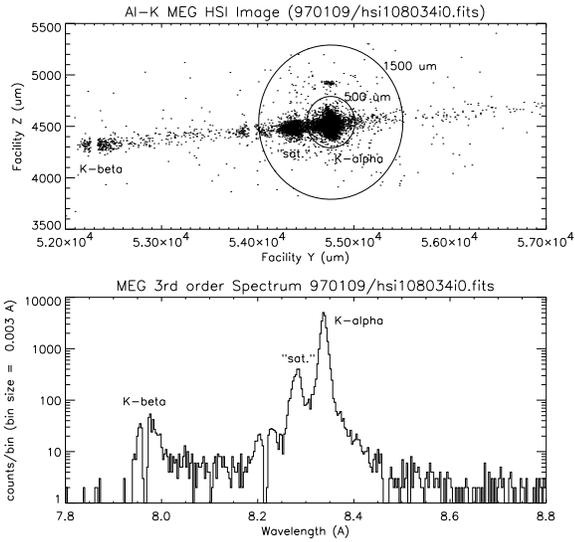}
\caption{XRCF image of 3rd-order MEG Al-K line (top)
and the resulting grating-produced spectrum (bottom.)
A strong ``satellite'' line is clearly visible near the
K-$\alpha$ peak.  This image was obtained with the
high speed imager, HSI, in the focal plane;
its instrumental gaps have not been
removed, {\it e.g.}, at the K-$\beta$ line.
Note the mis-aligned MEG grating outlier at Facility Z
of $\approx$4900; the extent in the Z-direction of the
main image is due to several more mis-aligned MEGs.
\label{fig:alkmeghsi}}
\end{figure}

Detailed images and measurements of HETG-diffracted X-ray lines were
made at XRCF to study the line response function, LRF \citep{marshall97}.
For example, Figure~\ref{fig:alkmeghsi} shows an image and resulting
spectrum
of the MEG $3^{\rm rd}$-order
diffracted Al-K line complex recorded at XRCF with the non-flight
High Speed Imager, HSI, micro-channel plate detector.
Measurements of various parameters related to the LRF are
described in the following paragraphs and summarized in
Table~\ref{tbl-1}.

\paragraph{Grating Angles}
  By measuring
the centroid of the diffracted images from HEG and MEG gratings, the
opening angle between HEG and MEG was measured to be very close to the 10
degree design value, see Table~\ref{tbl-1}.  

\paragraph{Grating Period and Rowland Spacing}
  Measurements using X-ray lines of known
wavelength were used to confirm the values
of the grating periods and measure the
HETG Rowland spacing at XRCF.
The ratio of HEG to MEG period determined from 
measurements agrees within a 100~ppm uncertainty
with the ratio expected based on the
lab-derived periods in Table~\ref{tbl-1}.
Adopting these periods and an Al-K energy of 1.4867~keV
the HETG Rowland spacing as-operated at XRCF was determined 
and is given in Table~\ref{tbl-1}.

\paragraph{LRF Core Measurements}
In order to see if the insertion of the HETG modified the
{\it zeroth-order image}, HSI exposures were taken in Al-K X-rays
of each shell of the HRMA illuminated in turn with no grating present.
With the HETG inserted images were obtained for the zeroth-order of
the MEG-only and HEG-only through each of four quadrants.  The HRMA
exposures for shells 1 and 3 (4 and 6) were then compared
with the combined MEG (HEG) zeroth-order quadrant images.
Comparing the projections of the two data sets binned to 
10~$\mu$m shows good agreement in shape, within 10-20\% in each bin, over 2 decades
of the PSF intensity, covering the spatial range $\pm 150\mu$m.
In particular, the inner core of the HRMA PSF at XRCF shows
a FWHM of $\approx 42\mu$m and insertion
of the HETG adds no more than an additional FWHM of $\approx 20\mu$m, i.e.,
at most increasing the FWHM from 42 to 46$\mu$m.

For {\it diffracted images}, precise measurements were made of the core of the
PSF by
using slit scans of the Mg-K 1.254 keV (9.887 \AA) line in the bright orders
$m=0,1,2$ for HEG and $m=0,1,3$ for MEG.
Scans were made along both the dispersion and
cross-dispersion directions using 10~$\mu$m and 80~$\mu$m wide slits in
front of proportional counter detectors.
To create simulated XRCF slit scan data,
a spectral model of the XRCF source was folded through a
MARX \citep{wise00} ray-trace simulation tailored to XRCF parameters
that affect the intrinsic LRF
(most importantly: finite source distance, finite source size, and
an additional 0.3~arc seconds of HRMA blur.)
The intrinsic FWHM of the line spectral model and the
period variation $dp/p$ values for each grating type (HEG, MEG)
were then adjusted in the simulation.
Good agreement with the XRCF data is obtained
when the core of the 
Mg-K line is modeled as a Gaussian with an $E/dE = 1800$, the
HEG gratings have a $dp/p$ value of 146~ppm~rms and the MEG
gratings have a $dp/p$ of 235~ppm~rms. 
These values are larger than
expected from the individual LR results and likely represent
slight additional distortions introduced during the facet-to-HESS
alignment and bonding
process; however, they are within our design goal of 250~ppm.

\paragraph{Mis-aligned MEG gratings}

\begin{figure}
\epsscale{1.0} 
\plotone{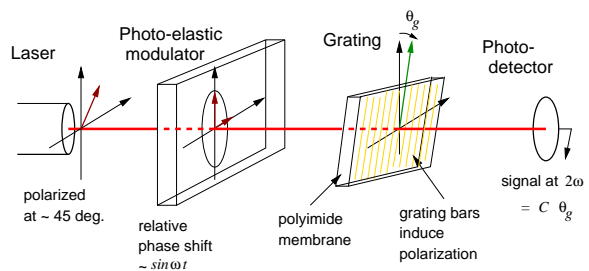}
\caption{
Schematic of the polarization alignment setup.  The intensity of the 
detected interference signal at $2\omega$ is proportional to $\theta_g$ for
small angles. Note that in this configuration the polyimide membrane
is between the modulator and analyzer (the grating bars), hence the
polarization/phase properties of the membrane can effect the measurement
result.
\label{fig:align_setup} }
\end{figure}

  As seen and noted in Figure~\ref{fig:alkmeghsi}, a small ``ghost
image'' is visible in the cross-dispersion direction
``above'' the diffracted Al-K line.
Analyzing similar images taken quadrant-by-quadrant
as well as a very (65.5~mm!) defocused image of the MEG 3rd
order which isolated the individual grating facet images \citep{marshall97},
we were able to demonstrate that this and
other ghost images closer to the main image were created by individual
MEG grating facets whose grating bars are ``rolled'' from the nominal
orientation.  In all, six of the 192 MEG facets have roll offsets
in the range of 3 to 25 arc minutes - greatly in excess of the
laboratory alignment system measurement results.
The individual facets were identified, for details
see \citet{CalReport}, and all came from fabrication Lot~\#7 ---
the only lot
which was produced with prototype fabrication tooling
during the membrane mounting step (Fig.~\ref{fig:fab_steps}g).  
It was subsequently demonstrated in the laboratory
(by Richard Elder) that inserting a stressed polyimide membrane between
the photo-elastic modulator and the grating, see Figure~\ref{fig:align_setup},
could create a shift in the alignment angle of order arc minutes
and which varied with applied stress.
Note in Figure~\ref{fig:align_setup} that
the polyimide layer of the facet being aligned is
in the optical path between the
polarization modulated alignment laser and the grating bars.
This clearly suggests that stress birefringence \citep[p.~703]{born80}
in the grating's polyimide membrane
introduces unintended bias offsets in the optical measurement of the
grating bar angles, effectively causing their mis-alignment by these
same bias angles.

\paragraph{Roll variations}
  Mg-K slit scans, described above, were also taken in the
cross-dispersion direction and provide a check on the roll
variations and alignment of the grating facets, the main contributor
to cross-dispersion blur beyond the mirror PSF and Rowland astigmatism.
Cross-dispersion distributions were input to the ray-trace simulations
and adjusted to agree with the data.  The resulting HEG and MEG roll
distributions each have an rms variation of 1.8 arc minutes.  The MEG
distribution is close to Gaussian while the HEG shows a clear
two-peaked distribution with the peaks separated by 3~arc~minutes.
These variations are larger than the 0.42~arc~minute~rms
value expected from the
polarization alignment laboratory tests.  The most likely cause is
polyimide membrane effects similar to those which produced the
mis-aligned MEGs but occurring at a lower level.  The mis-aligned
gratings and the roll distributions are explicitly modeled in MARX
simulations.

\begin{figure}
\epsscale{1.0} 
\plotone{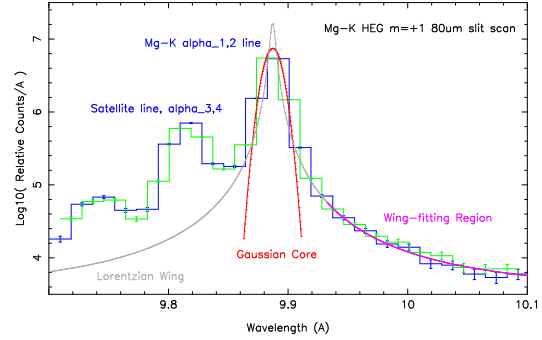}
\caption{Wings of the LRF.
At XRCF a focal-plane
proportional counter with an 80~$\mu$m by 500~$\mu$m aperture
was scanned across the dispersed Mg-K line image; two
interleaved scans (offset by 40~$\mu$m) of the HEG $m=+1$ order
are shown here.
To measure the wing level, the core of the LRF is
fit with a Gaussian and a region in the wings is fit
with a Lorentzian; these fit parameters are used to
quantify the wing level.
Most of the
wing flux seen here is due to the natural Lorentzian line shape
of the Mg-K line and not the HETGS instrument.
\label{fig:heg_wings}}
\end{figure}

\paragraph{Wings on the LRF}
  Wide-slit scans of the Mg-K line were used to set an upper limit to any
``wings'' on the LRF introduced by the HETG gratings.
The Mg-K PSF was scanned
by an 80~$\mu$m x 500~$\mu$m slit for the MEG and HEG mirror shell
sets separately.  These scans were
fit, using ISIS \citep{houck00} software,
by a Gaussian in the core and a Lorentzian
in the wings, as shown in Figure~\ref{fig:heg_wings}.
Quantitatively the wing level away from the Gaussian core can be expressed as:

\begin{equation}
L_W(\Delta\lambda) = A_G C_{\rm wing}/(\Delta\lambda)^2
\label{equ:winglevel}
\end{equation}

\noindent where $L_W$ is the measured
wing level in counts/\AA, $A_G$ is the area of the Gaussian core in counts,
and $\Delta\lambda = \lambda - \lambda_0$ is the distance from the line
center.  The strength of the wing is given by the value $C_{\rm wing}$
which has units of ``fraction/\AA~$\times$~\AA$^2$'' or (more opaquely)
just ``\AA''.
Using this formalism the observed wing levels were determined for the
HEG 1$^{\rm st}$ and 2$^{\rm nd}$ orders and the MEG 1$^{\rm st}$ and 3$^{\rm rd}$
orders, giving values of
8.6, 7.1, 12.2, and 7.8 $\times 10^{-4}$~\AA\ respectively.
Of these totals 5.6$\times 10^{-4}$ is due to
the intrinsic Lorentzian shape of the Mg-K line itself
 \citep[p.108]{agarwal91} with the reasonable value of a natural linewidth of 0.0035~\AA.
The remaining wing level can be largely explained as due to the
wings of the HRMA PSF itself:
because the LRF is essentially the HRMA PSF displaced along
the dispersion direction by the grating diffraction, 
wings of the mirror PSF directly
translate into wings on the grating LRF.
Subtracting these values we get an estimate of (upper limit on)
the contribution to the wing level by the HEG and MEG gratings per se,
as given in Table~\ref{tbl-1}.

\begin{figure}
\epsscale{1.0} 
\plotone{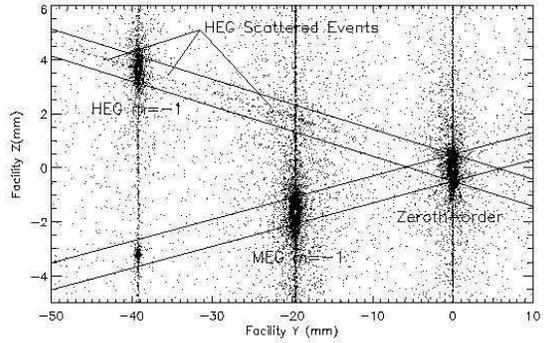}
\caption{Scattered events seen
in a monochromatic exposure at XRCF.  The
XRCF Double Crystal Monochromator was tuned to the Tungsten
1.3835~keV line for this HETG-ACIS-S exposure.
HEG scattered events are clearly visible concentrated
along the HEG
dispersion direction on either side of the HEG first-order and
near the one-half-order region.  In contrast, the MEG shows
few if any such events nor are such events seen near zeroth order.
\label{fig:heg_scatter}}
\end{figure}

\paragraph{Scatter beyond the LRF}
  Tests were also carried out at XRCF to search for any response
well outside of the discrete diffraction orders.
A high-flux, monochromatic line 
was created by tuning the Double Crystal Monochromator (DCM) to
the energy of a bright tungsten line from the rotating anode
X-ray source.  The HEG grating
set did show anomalous scattering of monochromatic
radiation \citep{marshall97}, in particular a small flux of
events with significant
deviations from the integer grating orders are seen
concentrated along the HEG dispersion direction,
Figure~\ref{fig:heg_scatter}. No such additional 
scattering is seen along the MEG dispersion direction.

The origin of this scattering was understood using an approximate model
of a grating with simple
rectangular grating bar geometry that incorporates 
spatially correlated deviations in the bar-parameters 
 \citep{davis98}, {\it i.e.}, there is Fourier
power in the grating structure at spatial periods other than the dominant
grating period and its harmonics.
Expressions for the correlations and the scattering
probability were derived and then fit to the experimental data. The
resulting fits, while not perfect, do reflect many of the salient
features of the data, confirming this as the mechanism for the scattering.  The
grating-bar correlations deduced from this model lead to a simple
physical picture of grating bar fluctuations where a small fraction of
the bars (0.5\%) have correlated deviations from their nominal geometry such
as a slight leaning of the bars to one side.  It is reasonable that the
HEG gratings, with their taller, narrower bars,
are more susceptible to such deviations than the MEG, which
does not show any measurable scatter.

In practice the scattered photons in HEG spectra are excluded from analysis
through order selection using the intrinsic energy resolution of ACIS:
the energy of the scattered photon is 
significantly different from the energy expected at its apparent
diffraction location.  This is not true for scattered events
that are close to the diffracted line image and they will make up a local
low-level pedestal to the HEG LRF.
However, the power scattered is small
compared to the main LRF peak, generally contributing
less than 0.01~\% of the main response into a three FWHM wide
region (0.036 \AA ) and less than 1~\% in total.

\paragraph{ACIS Rowland Geometry}
  An XRCF test was designed to verify the Rowland geometry of the
HETGS, in particular that all diffracted orders simultaneously come to best
focus in the dispersion-direction.  Data were taken with
each of four quadrants of the HRMA illuminated, allowing 
us to determine the amount of defocus for each of
the multiple orders imaged by the detector.  Because of
the astigmatic nature of the diffracted images, the axial location of
``best focus'' depends on which axis is being focused.
The results of this test \citep{stage98} were limited by the number of
events collected in the higher-orders; however it was concluded that
i) the astigmatic focal property was confirmed, ii) the HEG and MEG
focuses at XRCF differed by 0.32~mm as expected from HRMA modeling,
and iii) the ACIS detector was tilted by less than 10~arc~minutes 
about the Z-axis.

\subsection{Efficiency and Effective Area Measurements}\label{sec:efficEA}
A major objective of XRCF testing was to measure the efficiency  
of the fully assembled HEG and MEG grating sets and the
effective area of the full HRMA + HETG + ACIS system.
The HETGS effective area or ARF \citep{davis01a} for a given
grating set or {\it part}, indicating HEG or MEG, 
and diffraction order $m$ may be expressed in simplified form as:

\begin{equation}
A_{{\rm P},m}(\lambda) ~=~ M_{\rm P}(\lambda)~~ g_{{\rm P},m}(\lambda)~~ 
Q(\lambda,\vec{\sigma})~~
\end{equation}

\noindent where $\lambda$ denotes the dependence on
photon wavelength (or energy) and the
three contributing terms are the HRMA
effective area $M_{\rm P}(\lambda)$ for the relevant {\it part} ({\it e.g.},
MEG combines the area of HRMA shells 1 and 3),
the effective HETG grating efficiency
for the {\it part}-order
$g_{{\rm P},m}(\lambda)$, and the ACIS-S quantum detection
efficiency $Q(\lambda,\vec{\sigma})$.  
The $\vec{\sigma}$ parameter signifies 
a dependence on the focal-plane spatial location,
{\it e.g.}, at ``gap'' locations
between the individual CCD detector chips we have
$Q(\lambda,\vec{\sigma}_{\rm gap})=0$.  
Although not explicitly shown, this ACIS efficiency also depends on
other parameters, in particular CCD operating temperature and
event grade selection criteria.

\begin{figure}
\epsscale{1.0} 
\plotone{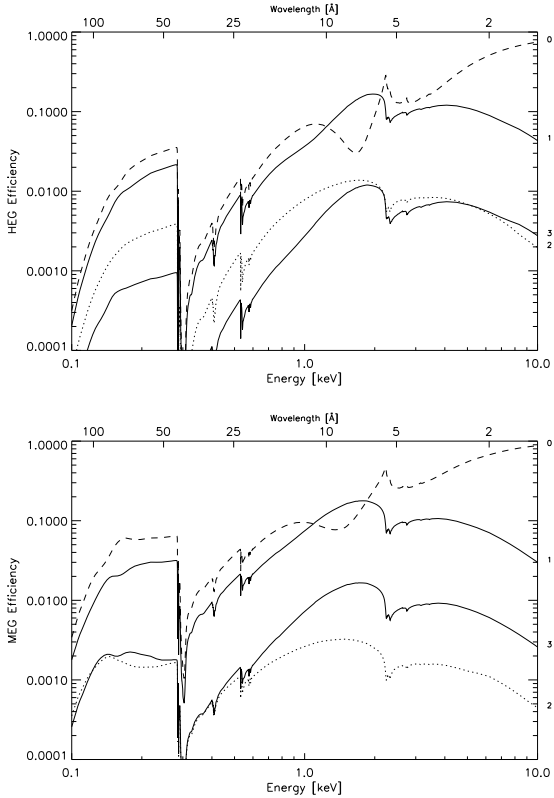}
\caption{HEG and MEG Effective Diffraction Efficiencies.
These single-sided efficiencies are the HEG and MEG
efficiencies averaged over the sets of facets and weighted
by the mirror shell areas.  The diffraction order is labeled by the
integers to the right of the plots; note that for the wider-barred
HEG the second order is generally higher than the third order
whereas the MEG shows the more expected suppression of the second order.
\label{fig:hegmeg_effics}}
\end{figure}

The grating effective efficiency $g_{{\rm P},m}(\lambda)$
is defined as:

\begin{equation}
g_{{\rm P=MEG[HEG]},m}(\lambda) = 
{{ \sum_{s=1,3[4,6]} M_s(\lambda) g_{s,m}(\lambda) }\over{
\sum_{s=1,3[4,6]} M_s(\lambda) }}
\end{equation}

\noindent where $s$ designates the HRMA shell (the numbering system is
a legacy from the original AXAF HRMA design which had 6 shells; 
shells 2 and 5 were deleted to save weight and cost).
The grating efficiency values $g_{s,m}(\lambda)$ are
the average of the facet efficiency models derived from X-GEF data
for all facets on shell~$s$
multiplied by a shell vignetting factor (primarily the fraction
of the beam not blocked by grating frames), Table~\ref{tbl-1}.
The values of these (single-sided) effective efficiencies are plotted
for zeroth through
third order in
Figure~\ref{fig:hegmeg_effics} for the HEG and MEG grating sets;
they are version ``N0004'' based on the laboratory measured
facet efficiencies and using our updated optical constants.

\paragraph{Diffraction Efficiency Measurements}
  In principle the diffraction efficiency of the HETG can be measured as
the ratio of the flux of a monochromatic beam diffracted into an order divided by the flux
seen when the HETG is removed from the X-ray beam, a ``grating-in over
grating-out'' measurement. If the same detectors are used in the
measurements then their properties cancel and the efficiency
can be measured with few systematic effects.  In early testing at XRCF
the HEG and MEG diffraction efficiencies were measured using
non-imaging detectors, a flow proportional counter and a solid state
detector \citep{dewey97,dewey98}.  The
detector's entrance aperture could be defined by a pinhole of
selectable size, typically 0.5 to 10~mm in diameter.

The main complication of these non-imaging measurements results from the
complexity of the source spectra and the limited energy resolution of
the detectors compared to that of the HETGS.  
The Electron Impact Point Source (EIPS)
was used to generate K and L lines of various elements, in particular
C, O, Fe, Ni, Cu, Mg, Al, Si, Mo, Ag, and Ti.  As Figure~\ref{fig:alkmeghsi}
shows, the ``line'' typically  consists of several closely spaced
lines.  For the ``grating-in'' dispersed measurement only some fraction
of these ``lines'' fall in the pinhole aperture and are detected
{\it e.g.}, consider the the $500~\mu$m aperture indicated in the figure.
In contrast, the ``grating out'' measurement includes all of the lines
and any local continuum within the energy resolution element of the
low-resolution detector.
In order to make a correction for this effect,
spectra at HETG resolution, similar to Figure~\ref{fig:alkmeghsi}, 
were collected for each X-ray line of interest and used
to calculate appropriate correction factors, ranging from a few percent to
a factor of two.

\begin{figure}
\epsscale{1.0} 
\plotone{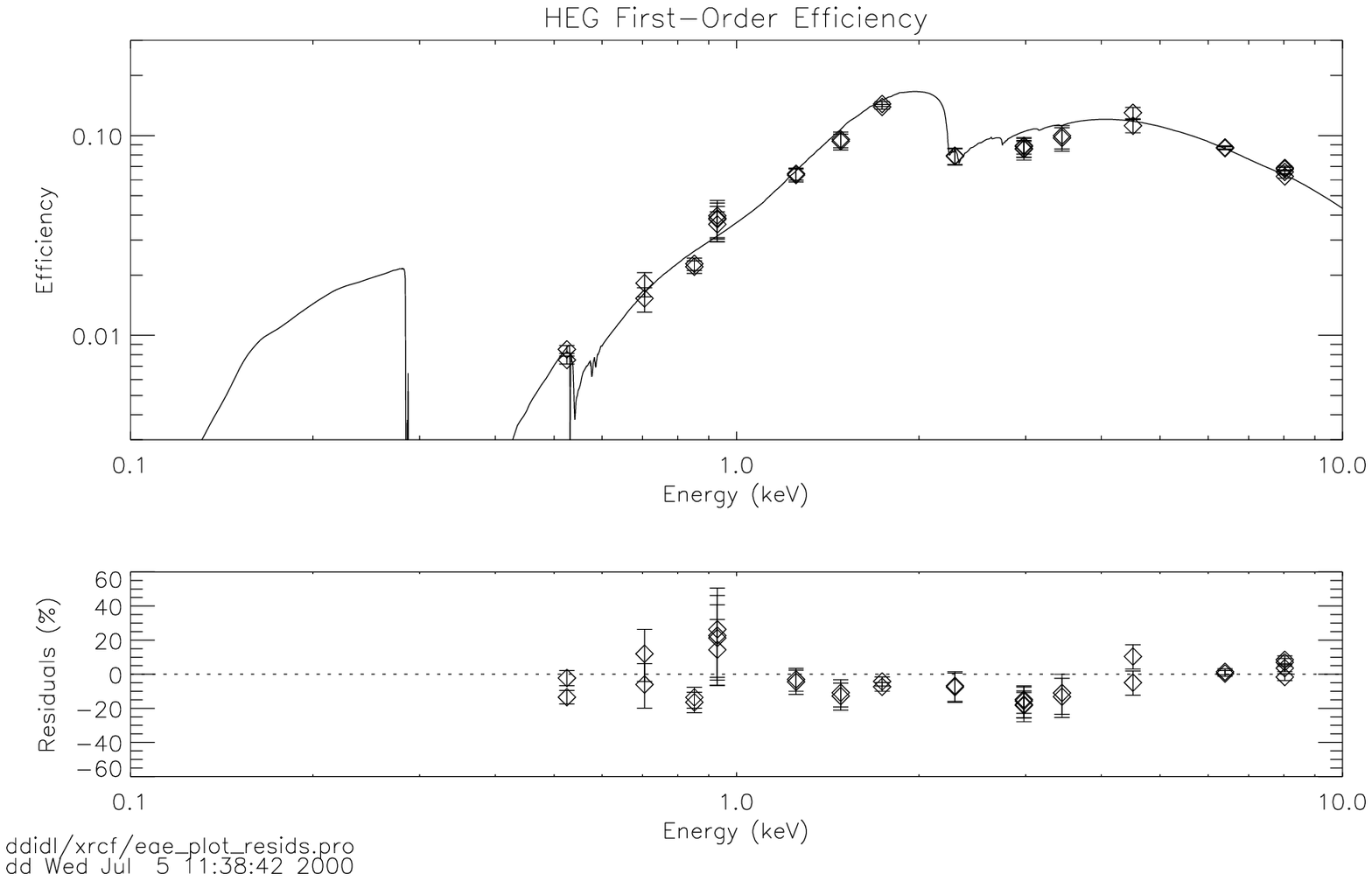}
\plotone{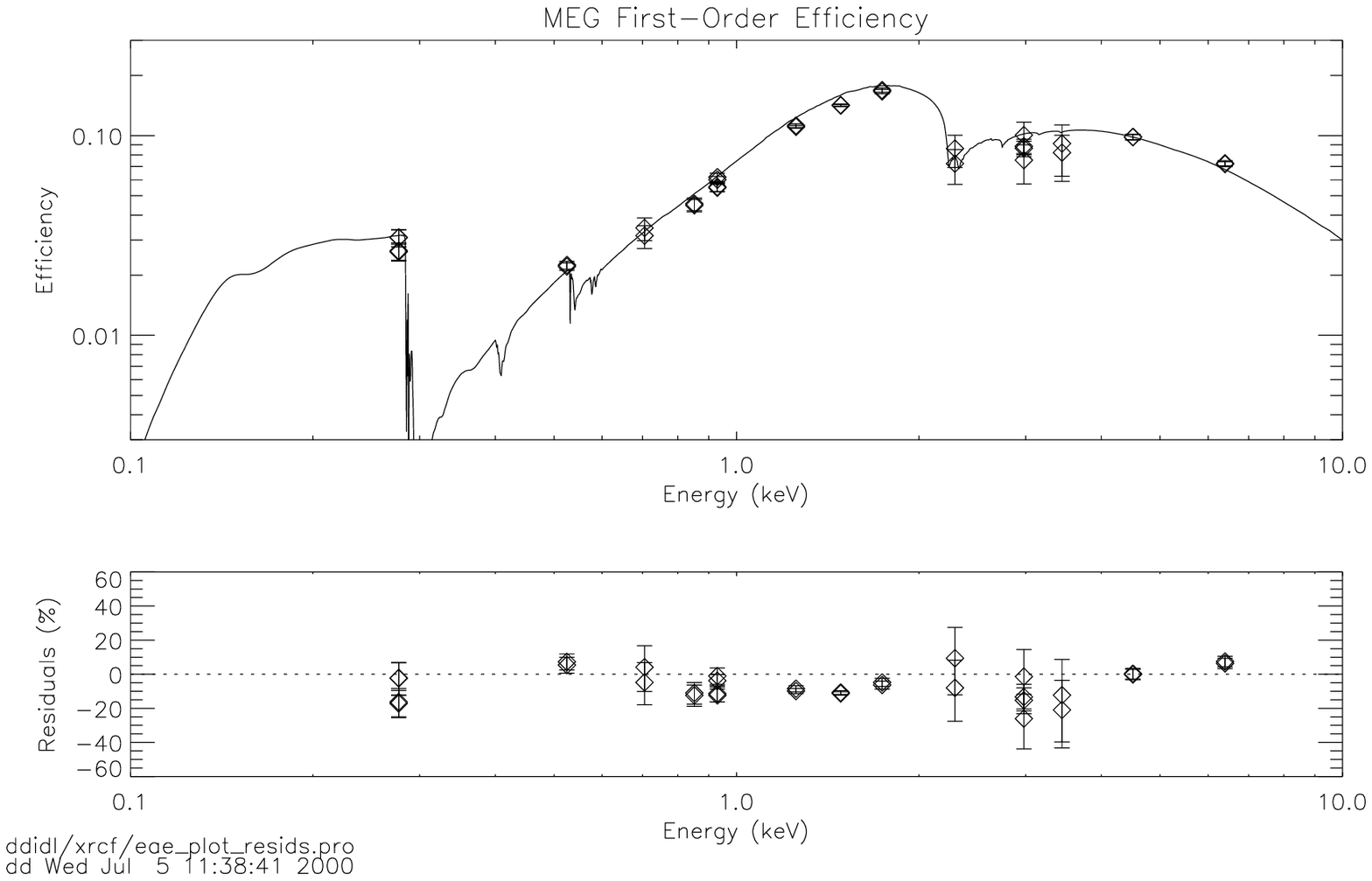}
\caption{HEG and MEG First-order, single-sided diffraction efficiency measurements
made with non-flight detectors at XRCF.
Error bars shown include systematic errors which arise from
corrections applied due to complex X-ray source line structure and the use
of non-imaging detectors, {\it e.g.}, for the high-energy
L-lines between 2 and 4~keV.  On the whole the measurements compare
well with the expected values (solid line.)
\label{fig:hegmeg1_effics}}
\end{figure}

The resulting XRCF efficiency measurements for the HEG and MEG
first orders are shown in
Figure~\ref{fig:hegmeg1_effics} along with the laboratory-based
values (solid lines, from Figure~\ref{fig:hegmeg_effics}.)
The error bars here, in addition to counting statistics,
include an estimate of the systematic uncertainty introduced 
in the correction process described.
These results
confirm the efficiency models derived from X-GEF measurements at the
10-20~\% level but also suggest possible systematic deviations.
Because these deviations are small we waited for flight data
before considering any corrections to the efficiency values.

\paragraph{Absolute Effective Area}
  Absolute effective area measurements were performed at the XRCF
with the flight ACIS detector, in particular the ACIS-S consisting
of a linear array of 6 CCD detector chips
designated S0 through S5 (or CCD\_ID = 4--9).
Devices at locations S1 and S3 are back illuminated (BI) CCDs
with improved low-energy response.  S3 is at the focal point, so
it detects zeroth order and is often used without the HETG inserted for
imaging observations; S1 is placed to detect the cosmically important
lines of ionized oxygen with enhanced efficiency.
Note that there are small gaps between the ACIS-S CCDs
with sizes determined by the actual chip focal plane locations.

\begin{figure}
\epsscale{1.0} 
\plotone{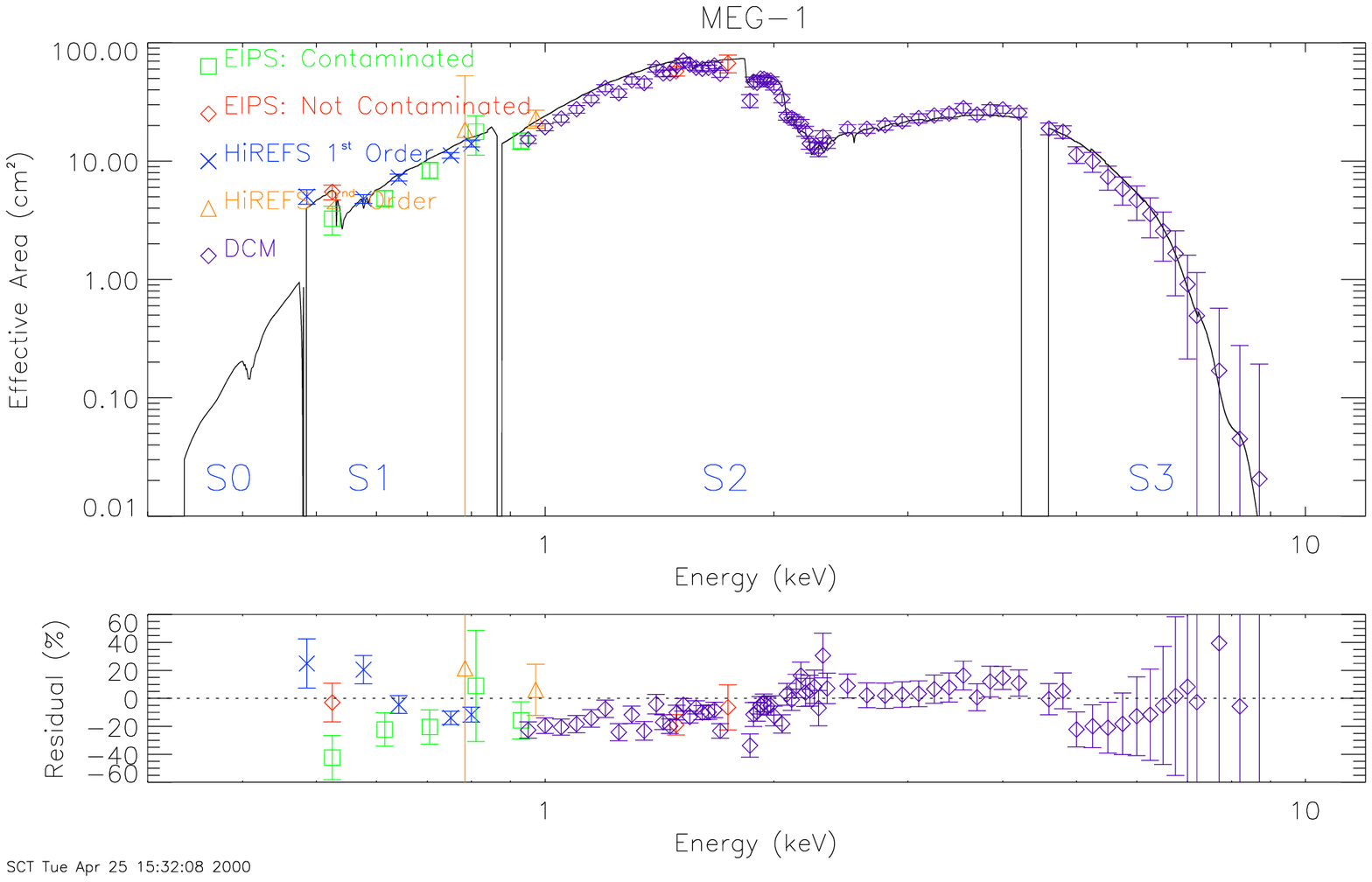}
\caption{Plots of measured and modeled
absolute effective area for the HETGS MEG {\it minus first} order
with residuals.
\label{fig:meg_areas}}
\end{figure}

\begin{figure}
\epsscale{1.0} 
\plotone{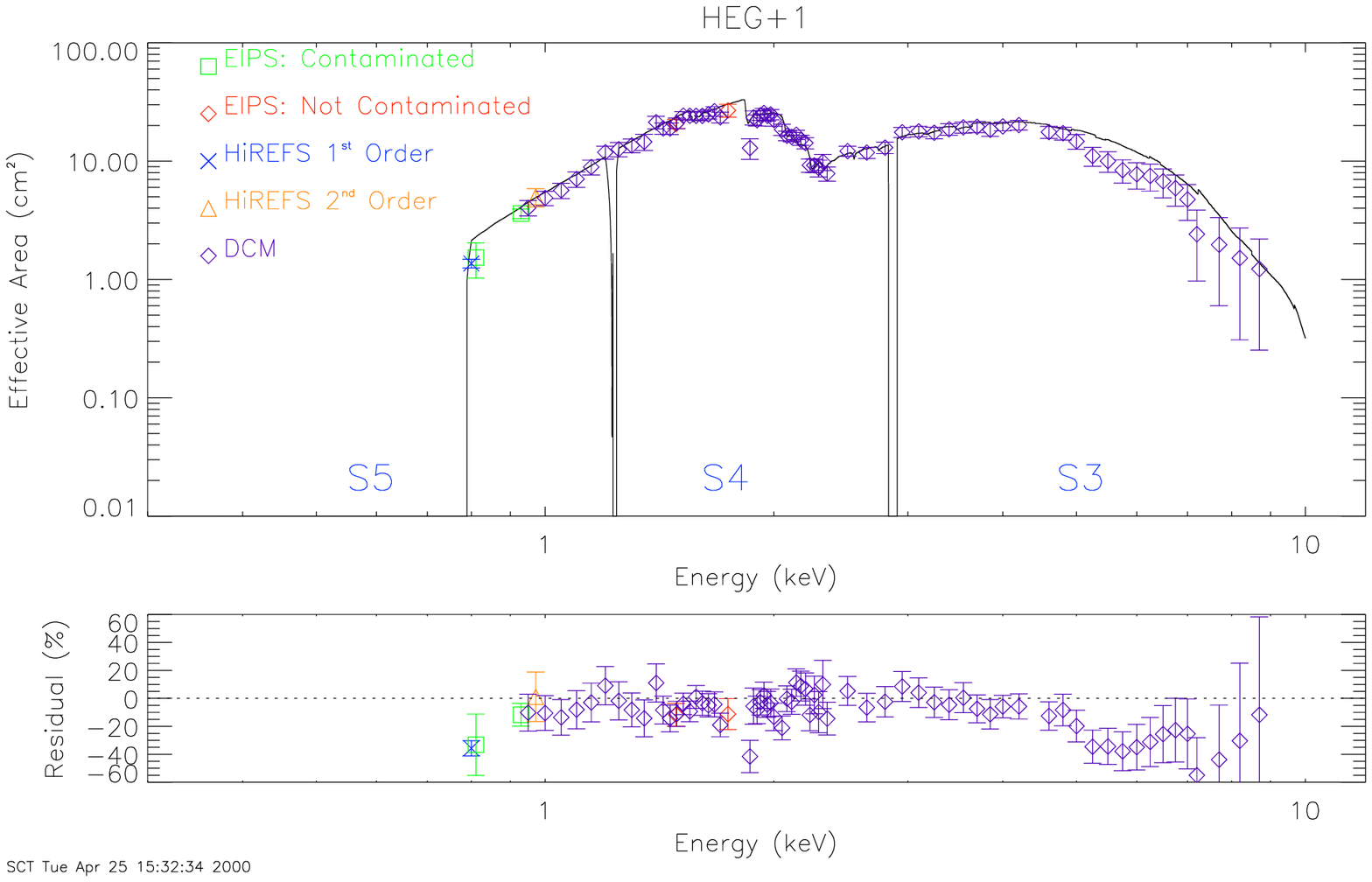}
\caption{Plots of measured and modeled
absolute effective area for the HETGS HEG {\it plus first} order
with residuals.
\label{fig:heg_areas}}
\end{figure}

The 1st-order HETGS effective area
can be divided 
into 5 regions where different physical mechanisms govern
the effective area of the system
(variously shown in
Figures~\ref{fig:multiv_effic},~\ref{fig:hegmeg_effics},
~\ref{fig:meg_areas}, and ~\ref{fig:heg_areas}):

\begin{itemize}
\item[ ] {\it below 1 keV}~--~Absorption by the polyimide membranes of
the gratings and the ACIS optical blocking filter and SiO layers
limit the effective area and introduce structure, with absorption
edges due to C, N, O, and Cr.
\item[ ] {\it 1-1.8 keV}~--~The phase effects of the
partially transparent 
gratings enhance the diffraction efficiency.
\item[ ] {\it 1.8-2.5 keV}~--~Edge structures are due to the Si (ACIS), 
the Ir (HRMA), and Au (grating).
\item[ ] {\it 2.5-5.5 keV}~--~Effective area is slowly varying, with some
low-amplitude Ir (HRMA) and Au (HETG) edge structure.  The efficiency is
also phase-enhanced in this region especially for the HEG.
\item[ ] {\it 5.5-10 keV}~--~The mirror reflectivity, grating efficiency,
and ACIS efficiency all decrease with increasing energy leading
to a progressively steepening decline.
\end{itemize}

The energy range from 0.48 to 8.7 keV was sampled at over 75 energies
using X-rays produced by three of the sources of the
X-Ray Source System \citep{jeffk95}.  The Double
Crystal Monochromator (DCM) provided dense coverage of the range 0.9 to
8.7 keV; the High Resolution Erect Field (grating) Spectrometer
(HiREFS) provided data points in the 0.48 to 0.8 keV range; and
X-ray lines from several targets of the Electron Impact Point Source (EIPS)
covered the range from 0.525 keV (O-K) to 1.74 keV (Si-K).

\begin{figure}
\epsscale{1.0} 
\plotone{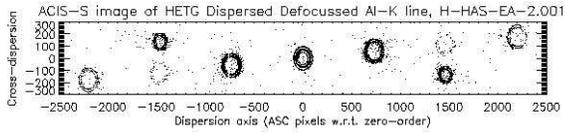}
\caption{ACIS-S defocused image of HETG dispersed
Al-K line at XRCF.  The rings from the four HRMA shells are visible
in the central, zeroth-order image.
The HEG and MEG dispersed orders are clearly
identified by the corresponding pairs of
HRMA shells in their images.  Only the $m=\pm 1$ order images are seen
for the HEG grating; with less dispersion the MEG orders $m = \pm 1,
\pm 2, \pm 3$ are all within the image.
\label{fig:acis_ea_image}}
\end{figure}

The absolute effective area was measured as the ratio of the focal plane
rate detected in a line to the line flux at the HRMA entrance aperture.
A set of four Beam Normalization Detectors (BND) were located around the
HRMA and served as the prime source of incident flux determination.
The ACIS detector was defocused by 5 to 40~mm to reduce
pileup caused when more than one photon arrives in a small region
of the detector during a single integration \citep{davis01b,davis03}
by spreading the detected events over a larger detector area as
seen in Figure~\ref{fig:acis_ea_image}.
A variety of analysis techniques and considerations were applied
to analyze these data \citep{schulz98}, chief among them for the monochromator data sets
were beam uniformity corrections
to the effective flux
based on extensive measurements and modeling carried out by 
the MSFC project science group \citep{swartz98}.
Other corrections were made for line deblending 
and ACIS pileup.  Uncertainties in the measurements were assigned
based on counting statistics and estimated systematics; typically
each measurement has an assigned uncertainty of order 10~\%.

\begin{figure}
\epsscale{1.0} 
\plotone{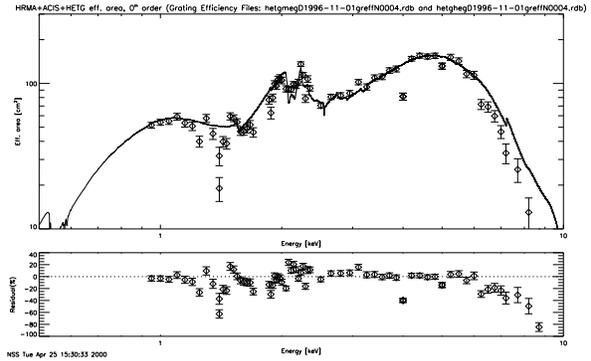}
\caption{Plot of measured and modeled
absolute effective area for the HETGS, HEG and MEG combined,
{\it zeroth} order with residuals.
\label{fig:hetg_zo_area}}
\end{figure}

Representative results are shown for
the MEG $m=-1$ in Figure~\ref{fig:meg_areas}, using ACIS chips
S0, S1, S2, and S3, and for the HEG $m=+1$
in Figure~\ref{fig:heg_areas}, detected on chips S3, S4, and S5.
Measurements of the HETG combined
zeroth order are shown in Figure~\ref{fig:hetg_zo_area}.
The data indicate that we are close to realizing our goal of a 10~\%
absolute effective area calibration for the first order effective
area.  The measurement-model residuals are seldom outside a $\pm$20~\%
range for both the HEG and MEG first-order areas.
Systematic variation of the residuals appear at a level of order
$-$20~\% in
the energy range below 1.3 keV; there is agreement
better than 10~\% in the 2.5 to 5 keV range,
The regions of greater
systematic variation, 1.3 to 2.5 keV and above 5 keV, are most likely
dominated by uncompensated DCM beam uniformity effects and ACIS 
pileup effects, respectively.

Effective area measurements for $|m| \ge 2$ were also carried out with
the flight focal plane detectors \citep{flanagan98} and
show agreement at the 20~\% level for HEG second and MEG third
orders.

\paragraph{Relative Effective Area}
  In order to probe for small scale spectral features in the HETGS
response we performed tests at XRCF using a very bright continuum
source \citep{marshall98}.
The Electron Impact Point Source (EIPS) was used with Cu and C
anodes and operated at high voltage and low current in order to
provide a bright continuum over a wide range of energies. The ACIS-S
was operated in a rapid read-out mode (``1x3'' continuous clocking mode)
to discriminate orders and to provide high throughput.

\begin{figure}
\epsscale{1.0} 
\plotone{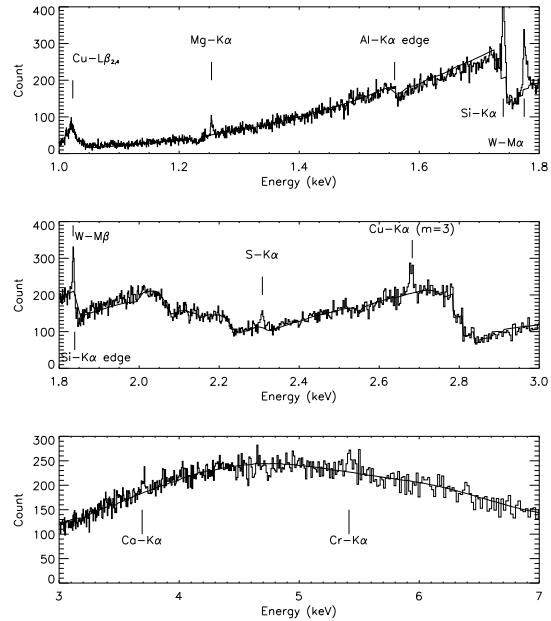}
\caption{HEG spectrum of the Cu continuum source.
These plots show an expanded view of the measured spectrum
(finely binned histogram) overlayed with a model based on
a smooth underlying source
spectrum folded through the mirror, HETG, and detector responses.
The well modeled detailed structure
of the ``bumps and wiggles'' in the
observed counts spectrum indicates an accurate relative effective area
calibration.
In addition to the expected
bright continuum, note the many weaker lines due to
contaminants in and on the source target.
\label{fig:mc_spectrum}}
\end{figure}

High-count spectra were created from the data and compared to a smooth
continuum model passed through the predicted HETGS effective area,
Figure~\ref{fig:mc_spectrum}.  Many spectral features are observed, 
including emission lines attributable
to the source spectrum. We find that models for the HETGS
effective area predict very well the
structures seen in the counts spectrum as well as the
observed fine structure near the
Au and Ir M edges where the response is most complex. Edges introduced
by the ACIS
quantum efficiency (QE) and the transmission
of its optical blocking filter are also visible,
the Si K and Al K edges respectively.
By comparing the positive and negative dispersion regions, we find no
significant efficiency asymmetry attributable to the gratings and we
can further infer that the QEs of all the ACIS-S frontside illuminated
(FI) chips are consistent to $\pm$10~\%.

\section{Five Years of Flight Operation}

\subsection{Flight Data Examples}

The first flight data from the HETG
were obtained on August 28, 1999 while pointing
at the active coronal binary star Capella.
Subsequent observations of this and other bright sources
provided in-flight verification and calibration.  The instrument
performance in orbit is very close to that measured and modeled on the ground.
A recent summary of \chandra's initial years is given by
\citet{weisskopf04} and \citet{paerelskahn03} review some
aspects of high resolution spectroscopy performed with
Chandra and XMM-Newton.

\begin{figure} 
\plotone{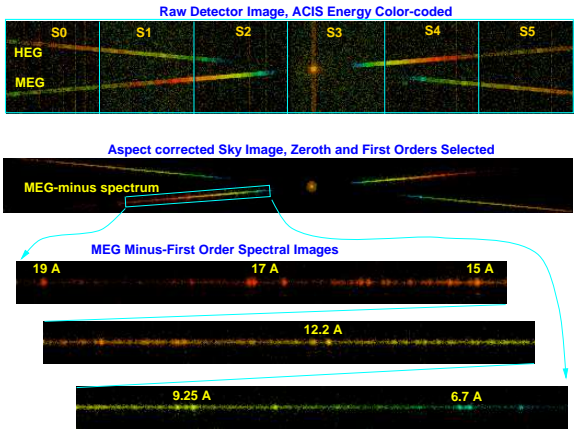}
\caption{Flight HETGS observation of Capella. In the top image the
HETG spreads the HRMA-focused X-rays into a shallow ``X'' pattern
on the ACIS-S detector and the spacecraft
dither broadens the image.
In the middle image the zeroth-order and dispersed images
are sharper because of aspect correction. (Note that this sky
coordinates image has been rotated and flipped to match the 
detector image orientation.)
At bottom
a wealth of spectral information is seen in the expanded MEG
minus-first order spectral image showing bright emission lines.
\label{fig:x_pattern}}
\end{figure}

Figure~\ref{fig:x_pattern} shows 26 ks of data from Capella. 
The top panel shows an image of detected events on the
ACIS-S detector with the image color indicating the ACIS-determined
X-ray energy.  In this detector coordinate image, the
features are broad due to the nominal dither motion in which
the spacecraft pointing is intentionally ``dithered''
to average over small-scale detector non-uniformities.  The ACIS-S chips are numbered
S0 to S5 from left to right, with the aim point in S3 where the bright
undispersed image is visible and includes a vertical frame-transfer
streak.
HETG-diffracted photons are visible forming a shallow ``X'' pattern.
The middle panel shows an image after the data have been aspect
corrected and data selections applied to include only valid zeroth and
first order events. The lower set of panels shows an expanded view of
the MEG, $m=-1$ spectrum with emission lines clearly
visible.  The observed lines and instrument throughput are
roughly as expected \citep{capella00}.

\begin{figure}
\epsscale{1.0} 
\plotone{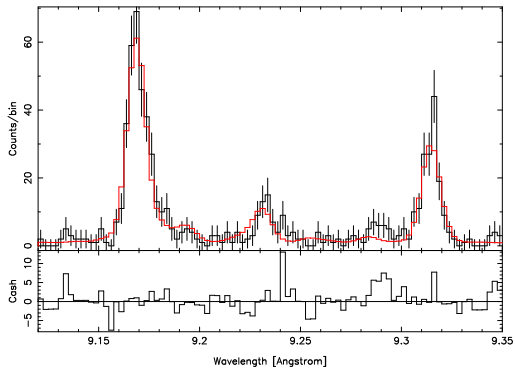}
\caption{Example of HEG resolving power and modeled LRF.
Shown is a closeup of the He-like Mg line complex near
9.25~\AA~(1.34 keV)
as seen by the HEG in 40.5~ks of Capella data (histogram.)
A model folded through the HEG instrumental
response is also shown (red) and has
a FWHM of order 11 m\AA~(1.6 eV) for a resolving power of $\approx 850$.
\label{fig:lrf_quality} }
\end{figure}

As a demonstration of the high resolving power of the HEG grating,
a closeup of the 9.12~\AA\ to 9.35~\AA\ 
spectral region of a Capella observation is shown in
Figure~\ref{fig:lrf_quality}. The three main lines seen here are
from $n=2$~to~$n=1$ transitions of the He-like Mg$^{+10}$ ion,
designated ``Mg \uppercase\expandafter{\romannumeral11}'';
a resolving power of $\approx 850$ is being achieved here
with a FWHM of $\approx 1.6$~eV.

\subsection{Flight Instrument Issues}

Since the HETGS is a composite system
of the HRMA, HETG, ACIS, and spacecraft systems, the HETGS
flight performance is sensitive to the properties of all of
these systems.  The various flight issues that have arisen
in the past 5 years are summarized here by component and
their effect on the HETGS performance is mentioned.  A complete
account of these issues is beyond the scope of this paper;
further details of the in-flight HETG calibration are presented in \citet{marshall04}.
See also documents and references
from the \chandra\ X-ray Center \citep{pogv7} which also
archives and maintains
specific calibration values and files
in the \chandra\ Calibration Database, CALDB, along with extensive
release notes.

\subsubsection{HRMA issues}

  The HRMA \citep{leon97} is the heart of the observatory and 
has maintained a crisp, stable focus for five years;  the commanded
focus setting has remained the same for five years.
The HETG resolving power has remained stable as well indicating
stability of the grating facets and overall assembly.  Details of the
HRMA PSF in the wings are still being worked but this has
minimal effect on the HETG LRF/RMF in practical application.

\begin{figure}
\epsscale{1.0}
\plotone{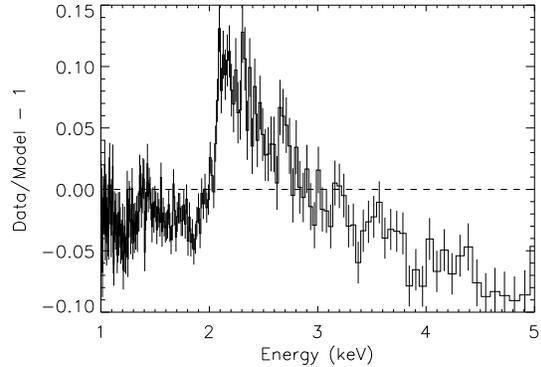}
\caption{Deviations at the Ir-edge seen with HETG.
The HETG counts spectrum clearly shows the structure
of the residual between data and model at the Ir edge and extending
to higher energies.  Note that the full range plotted
here is only $-10$\% to $+15$\%.  
This structure can be reasonably
explained as the effect of a 20~\AA\ hydrocarbon
contaminant layer on the HRMA.
\label{fig:ir_edge}}
\end{figure}

  The only issue arising in flight related to the HRMA
is seen as a slight step-increase (15~\%)
in effective area in the region near the Ir M-edge - seen clearly
in HETGS spectra.
A model based on the
reflection effect of a thin contamination layer on the HRMA optical surfaces
agrees reasonably with the deviations seen,
Figure~\ref{fig:ir_edge}, implying a hydrocarbon layer thickness of 20$\pm$5~\AA.
At present there is no evidence that the layer thickness is changing significantly with time;
detailed modeling and updates to the calibration products are in progress.

\subsubsection{ACIS issues}

  In the first flight data sets slight wavelength differences were seen
between the plus and minus orders indicating a few pixel error in our knowledge of
the relative locations of the 
ACIS-S CCDs.  The ACIS geometry values were adjusted for this in 1999
to an accuracy of $\sim 0.5$~pixels.  As a by-product of our HETG LSF 
work, see below,
these values have been updated a second time to an accuracy of 
$\sim 0.2$~pixels and they show a stability at this level
over the first five years of flight.

  The ACIS pixel size as-fabricated
was precisely quoted as 24.000~$\mu$m and this value was used
for initial flight data analysis.  However it was later realized that
this value was for room temperature - at the flight operating temperatures of
order $-120$~C the pixel size was determined to be 23.987~$\mu$m
and this value has been incorporated into the analysis software, etc.

\begin{figure}
\epsscale{1.0}
\plotone{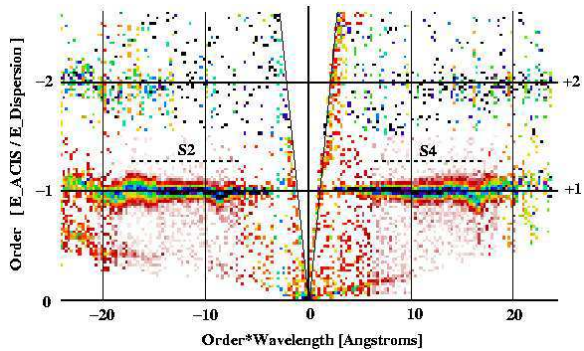}
\caption{Order separation with the ACIS-S.  The intensity of MEG events
extracted
from a Capella observation (obsid 3674) are indicated by color (from
red to blue/black) in order, $m$, vs.dispersion, $m\lambda$,
space.  The $x$-axis is equivalent to the dispersion location of the events
and the $y$-axis is the CCD determined energy expressed
as the ``order'': $m = E_{\rm ACIS} / E_{\rm Dispersion}$.
The regions readout by FI CCDs S2 and S4 are indicated; even with their
degraded resolution the order selection can be done with high efficiency.
\label{fig:order_sel}}
\end{figure}

  Order separation is performed using the intrinsic energy resolution of
the ACIS CCDs, as demonstrated in Figure~\ref{fig:order_sel} for
a Capella observation.
ACIS suffered some radiation damage early in the
mission which degraded the energy resolution of the front-illuminated, FI,
chips \citep{townsley00}; relevant to the HETGS are FI chips: S0, S2, S4, and S5.
However, as seen in
the figure, the resolution of those CCDs is
still sufficient to permit clean separation of the HETG orders.
The CXC pipeline software generally includes 95~\% of the first-order
events in its order selection; the exact fraction depending on CCD
and energy is calculated and included in the creation of HETGS ARF
response files.

  The ACIS detector suffers from pileup \citep{davis01b} and this was
expected in the bright zeroth-order image.  However, pileup also 
shows up in the dispersed spectra of bright sources and/or bright
lines; algorithms have been developed to ameliorate this
pileup \citep{davis03}.

  Early in the mission there were indications in
LETG-ACIS data that the C-K edge
of the ACIS optical blocking filter, OBF, was deeper than predicted.
Later it was realized that a contaminant was building up on the
ACIS OBF and hence the effective ACIS QE was decreasing
\citep{marshall04contam}.
The main spectral, temporal, and spatial effects of this
contaminant have now been incorporated into \chandra\ responses;
the composition and properties of the contaminant 
are the focus of continued measurement and modeling.

  Comparison of the plus and minus orders of the HETGS lead to a
measurement of a discrepancy of the QEs of the ACIS FI CCDs compared
to ACIS back illuminated CCDs (S1, S3.)
This issue is recently resolved into two components:
the FI devices suffer from cosmic-ray dead time effects of order
4~\%\ and the BI QEs are actually somewhat larger than initially calibrated.
The BI QEs were updated in August 2004 (CALDB 2.28) and
are thus now included in HETGS ARFs.

\subsubsection{HETG issues}

All the essential parameters for the HETG in flight are the same or
consistent with the ground values. Some notable quantities are discussed
below.

\noindent
{\it Clocking Angle:}  The flight angles of the HEG and MEG dispersion axes 
measured on the ACIS-S are given in Table~\ref{tbl-1}; these values are in
agreement with the XRCF-measured values.

\noindent
{\it Rowland Geometry and Spacing:} An accurate account of the Rowland
	geometry and spacing is crucial to achieve the best focus of the
	dispersed spectra on the detector. The Rowland geometry of the
	HETG was demonstrated during initial plate-focus tests:
	dispersed line images from a range of wavelengths came to their
	best dispersion-direction focus at a common detector focus
	value, which agreed with the ACIS-S3 best-imaging focus value
	within 50 $\mu$m.  The spacing of the HETG from the focal plane, the
	Rowland spacing, appeared initially to be off from
	the expected value until the ACIS pixel size change with
	temperature was included (previous section).  Currently the
	HETG Rowland spacing in flight, given in Table~1, is
	the value produced by ground installation metrology.

\begin{figure}
\epsscale{1.05}
\plotone{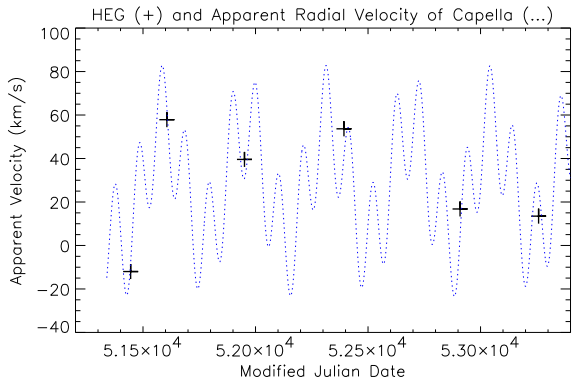}
\caption{Stability of the HETG wavelength scale over 5 years.
The measured  line centroid variation from Capella observations ($+$'s)
shows agreement and stability with the predicted
Capella Doppler motion at the 10~km/s level.
\label{fig:capella_period}}
\end{figure}

\noindent
{\it Grating Period:} The accuracy of the HETG-measured wavelength depends
	strongly on the assigned grating period. For the first years of use, the
	grating periods of the HEG and MEG were set to the laboratory measurement
	results.  However, recent analysis of Capella data over 5
	years shows the MEG-derived line centroid to be off by 40.2
	$\pm$5 km~s$^{-1}$ compared to the, apparently accurate,
HEG values, see Figure~\ref{fig:capella_period}.
	Hence the MEG period has recently (CALDB version 3.0.1, February 2005)
been set to 4001.95~\AA, which makes
	the MEG/HEG line centroids mutually consistent.  Note that the
	stability of the wavelength scale is good to 10 km~s$^{-1}$ or 30
	ppm over 5 years.

\noindent
{\it Dispersion and Cross-Dispersion Profiles:} Recently the HETG line
	response functions (LRFs) have been modeled as a linear
	combination of two Gaussians and two Lorentzians to encode
	improved fidelity with the latest results of MARX ray-trace
	models \citep{marshall04}; these LRF products are available
        in Chandra CALDB versions 
	2.27 and higher.  A Capella in-flight calibration data set (ObsID
	1103) has been used to verify the quality of the LRFs. Using a
	multi-temperature APED thermal model \citep{brickhouse96},
	the He-like line complex of
	Mg~XI has been fitted with the grating line response
	functions. Thermal broadening of Mg~XI species has been
	taken into account in order to measure its line width
	properly. Figure~\ref{fig:lrf_quality} shows the result of this model fitting.
	The derived line widths are
	essentially in agreement with having no excess broadening, as
	expected for static coronal emission.
	Note that the wings of the grating line response
	functions, two orders of magnitude below the peak when a few FWHMs
away, are generally well below the actual continuum and
	pseudo-continuum levels in celestial sources,
	and so flight dispersed data does not help calibrate the wings of
	LRF in general.

	In the cross-dispersion direction, we parameterize the 
	fraction of energy ``encircled'' in an arbitrary rectangular 
	region (the encircled energy fraction, or EEF) and include
this in the analysis software.
The calibration values encoded into LRFs 
	are generally consistent with observations.
An uncertainty 
	of 1 -- 3\% may be introduced by the EEF term, though other 
	quantitative uncertainties (e.g., HRMA effective area) are
comparable or
	greater at this point in HETGS calibration.

\noindent
{\it HETG Efficiency Calibration:}
       After five years of flight operation the efficiencies of
Figure~\ref{fig:hegmeg_effics}, based on the facet laboratory
measurements, have not been adjusted and are still used to create ARFs
for flight observations.  Likewise no additional features or edges have been
ascribed to the HETG instrument response beyond what is in these
calibration files.  Comparing HEG and MEG spectra of bright continuum
sources, there is data to support making a {\it relative} correction
of the HEG and MEG efficiencies to bring their measured fluxes into agreement.
This relative efficiency correction is small,
in the range 0~\% to -7~\% if applied to the MEG,
and varies smoothly in the 2~\AA\ to 15~\AA\ range.
This final ``dotting the i'' of HETG calibration is nearing
completion and will be included in upcoming CXC calibration files.

\subsection{Discussion}

As the minimal effect on the HETGS performance of the various flight
issues described above indicates, the HETGS has and is performing
essentially as designed yielding high-resolution spectra
of a broad range of astrophysical sources.
Some calibration issues are still being addressed, but
these are at the $\sim 10$~\%
level in the ARF and the fractions-of-a-pixel level in the grating LRF/RMF.

\begin{figure}
\epsscale{1.05} 
\plotone{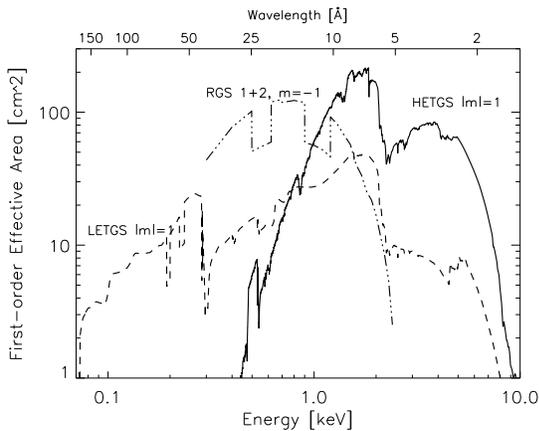}
\caption{
Effective areas for the \chandra\ HETGS (HEG+MEG) and LETG gratings and the
Reflection Grating Spectrometers (RGS 1+2) on XMM Newton.
The combined first-order areas are plotted for each instrument.
\label{fig:compare_areas} }
\end{figure}

\begin{figure}
\epsscale{1.05} 
\plotone{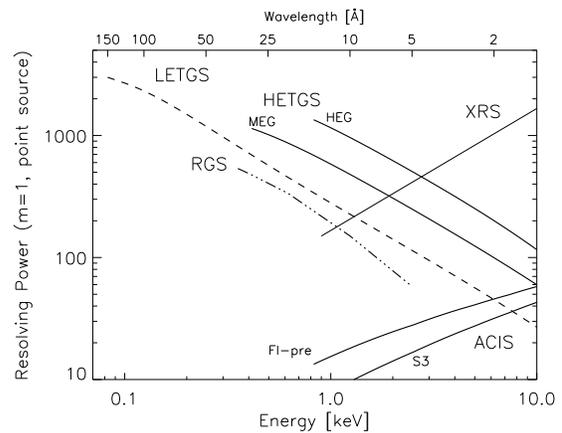}
\caption{
Resolving power in first order
for the \chandra\ gratings, HEG, MEG, and LETG, and
the RGS on XMM Newton.
For reference, representative resolving powers of the non-dispersive
\chandra\ ACIS FI (pre-launch) and BI (S3) detectors
and a micro-calorimeter X-ray Spectrometer, XRS (FWHM=6~eV),
are shown as well.
\label{fig:compare_respower} }
\end{figure}

It is useful to put the \chandra\ gratings' performance in
perspective with each other and with the Reflection Grating
Spectrometer, RGS, on XMM-Newton.
The effective areas
for these grating instruments are shown in
Figure~\ref{fig:compare_areas}; note
the complementary nature of the instruments in various
wavelength ranges.  The advance in 
resolving power these dispersive instruments have provided is 
clearly seen in Figure~\ref{fig:compare_respower} in 
comparison to the ACIS CCD resolving power values
shown there as well.  A line for the resolving power
of a 6~eV FWHM micro-calorimeter (e.g., Astro-E2) is
plotted as well showing the uniqueness of the grating
instruments in the range below 2~keV.

The high resolution and broad bandpass of the HETGS have made it the
instrument of choice for many observers.  In the first five years of
Chandra operation, the HETGS was used in over four hundred
observations totaling approximately 20~Ms of exposure time.  This
represents about 17~\% of Chandra's total observation time for that
period.  As is typical of spectroscopy at other wavelengths, HETGS
observations tend to be long, ranging from tens of ks for bright
Galactic sources to hundreds of ks for active galactic nuclei (AGN).
A review of the results of these observations is outside the scope of
this paper; some examples can be found in \citet{weisskopf02,weisskopf04}
and \citet{paerelskahn03}.

\acknowledgments

\paragraph{Acknowledgments}
  The HETG is the product of nearly two decades of research and development
in the MIT Center for Space Research, now the MIT Kavli Institute (MKI),
to adapt the 
techniques of micro- and 
nano-structures fabrication developed primarily for micro-electronics
to the needs of X-ray astronomy.  The initial concept emerged from 
nanostructures research conducted by one of us (H.I.S. and collaborators) in
the MIT Nano Structures Laboratory of the
Research Laboratory of Electronics 
(and previously at MIT Lincoln Laboratory).
We thank Al Levine, Pete Tappan, and Bill Mayer for initial
HETG design evaluations.

The Space Nanotechnology Laboratory was established in CSR under the
leadership
of HETG Fabrication Scientist M.L.S. to advance the 
technology and then execute the fabrication of 
the flight hardware.
Significant processing support was provided by the MIT
Microsystems Technology Laboratory.
Fabrication engineering support was provided by Rich Aucoin, Bob Fleming,
Pat Hindle, and Dave Breslau.
Facet fabrication was carried out by Jeanne Porter, Bob Sission,
Roger Millen, and Jane Prentiss.

D.D., Instrument Scientist, M.M.,  Systems
Engineer, and K.A.F. led the overall design and ground calibration activities.
J.E.D. provided efficiency modeling algorithms and software.
Design and test engineering support was provided by Chris Pak, Len Bordzol,
Richard Elder, 
Don Humphries, and Ed Warren. Facet testing was carried out by
Mike Enright and
Bob Laliberte who assembled the flight HETG.

H.L.M. carried out much of the XRCF test planning and analyses.
M.W. and D.P.H. provided modeling and analysis support.
N.S.S. analyzed key XRCF data as did Sara Ann Taylor and Michael Stage.
Many of the authors were involved in flight calibration; 
K.I. carried out detailed flight LSF work.
Finally, focus and progress were maintained
under the overall leadership of C.R.C., Instrument Principal Investigator,
supported by T.L.M. as
Project Scientist through much of the design and development phase
and E.B.G. as Project Manager \citep{galton03}.

The HETG team acknowledges conspicuous and inconspicuous
support from our colleagues at the
CSR and from the many groups involved in
the \chandra\ project, specifically
MSFC Project Science, Smithsonian Astrophysical Observatory,
TRW, and Eastman Kodak.  We thank
John Kramar and colleagues at NIST for LR period calibration.

Finally, we thank our fellow citizens: this work was supported by
the National Aeronautics and Space Administration under
contracts NAS8-38249 and NAS8-01129 through the Marshall Space
Flight Center.

Facilities: \facility{CXO(HETG)}

\appendix

\section{Multi-vertex Efficiency Equations}
\label{sec:effic_append}

Assuming the validity of scalar diffraction theory and ignoring
reflection and refraction, the $m$th order
grating efficiency for a perfect diffraction grating is
$|F_m(k)|^2$, where the structure factor $F_m(k)$ is given by
\begin{equation}
  \label{eq:fm1}
  F_m(k) = \frac{1}{p} \int_0^p \d x\; e^{i2\pi m x/p + i\phi(k,x)}.
\end{equation}
Here $p$ is the grating period, $k = 2\pi/\lambda$ is the wave-number,
and $\phi(k,x)$ is a phase shift introduced by the grating bars.  The
phase shift is a function of energy or wave-number $k$ and also depends
upon the grating bar shape according to
\begin{equation}
   \phi(k,x) = -k [ \delta(k) - i\beta(k) ] z(x),
\end{equation}
where $\delta$ and $\beta$ are energy dependent functions related to
the dielectric constant of the grating bars.  The function $z(x)$
represents the path length of the photon as it passes through a
grating bar; it is sometimes called, rather loosely, the ``grating bar
shape'', and more rigorously, the ``path-length function''.

It is preferable to work with the unit less quantity $\xi=x/p$ and to
parameterize the path-length function in terms of it.  For simplicity,
we represent $z(\xi)$ as a piece-wise sum of $N$ line segments,
{\it i.e.},
\begin{equation}
  z(\xi) = \sum_{j=0}^{N-1} (a_j + b_j \xi) B(\xi_{j}\le\xi\le\xi_{j+1}),
\end{equation}
where $B(X)$ is the boxcar function defined to be $1$ if $X$ is true,
or zero otherwise.  By demanding that the path-length function be
continuous, it is easy to see that the coefficients $a_j$ and $b_j$
are given by
\begin{eqnarray}
  a_j &=& \frac{z_j\xi_{j+1} - z_{j+1}\xi_j}{\xi_{j+1} - \xi_{j}},\\
  b_j &=& \frac{z_{j+1}-z_j}{\xi_{j+1}-\xi_j},
\end{eqnarray}
where $z_j = z(\xi_j)$.
For obvious reasons, we require $z_j\ge 0$, and that the set of points
$\{\xi_j\}$ be ordered according to
\begin{equation}
   0 = \xi_0 \le \xi_1 \le \cdots \le \xi_{N-1} \le \xi_N = 1.
\end{equation}

The most redeeming feature of this particular parameterization of the
path-length function is that the integral appearing in Equation
\ref{eq:fm1} may be readily evaluated with the result
\begin{equation}
 F_m(k) = i \sum_{j=0}^{N-1}
    e^{-i\kappa a_j} \frac{e^{-i\xi_{j+1}(\kappa b_j+2\pi m)}
                     - e^{-i\xi_j(\kappa b_j + 2\pi m)}}{\kappa b_j +
                     2\pi m},
\end{equation}
where
\begin{equation}
   \kappa = k [ \delta(k) - i\beta(k) ]
\end{equation}
is complex.  Although one may carry out the algebraic
evaluation of $|F_m(k)|^2$
using the above expression, it is very tedious and the result is not
particularly illuminating.  Moreover, it is computationally more
efficient to evaluate the above sums numerically using complex
arithmetic and then compute $|F_m(k)|^2$ by multiplying by the complex
conjugate.

\clearpage

\section{Error Budget for Faceted-Rowland Design}
\label{sec:rowland_append}

The response of the HETGS can be crudely yet usefully characterized
by the location and FWHM of the LRF core in both the dispersion and
cross-dispersion directions.  The ``resolving power'' of the spectrometer
is given by $E/dE = y'/dy'$ where $y'$ is the diffraction distance
and $dy'$ is the FWHM of the resulting image projected along the
dispersion axis.
The design of the HETG involved the use of an error budget to
assess and rss-sum the
various contributions to the ``$dy'$'' term of the
resolving power and the corresponding ``$dz'$'' term in the
cross-dispersion.  This error budget was
useful for studying the dependence of the resolving power
on the variation of individual error terms.  The error budget results
were verified by performing simplified ray-traces of single and
multiple facets.

The error budget presented in Table~\ref{tab:error_budget} includes
all of the important error terms for the flight HETGS resolving power
and cross-dispersion blur.
Note that the finite facet error term, equation~\ref{equ:ff_ppblur},
is {\it not} included here because it is quite small for the HETG design.
For compactness the error equations are referenced
in the table and given, with discussion, in the following text.

\paragraph{Optics PSF Blur} \label{sec:hrma_psf}
  If the optic produces a roughly symmetric Gaussian-like PSF
with an rms diameter of $D_{\rm PSF}$ arc seconds, then the
Gaussian sigma of the 1-D projection of the PSF is given, in
units of mm in the focal plane, by:
\begin{equation}
  \sigma_{y'} = \sigma_{z'} = \sigma_{H} = 
	{\frac{1}{2}}{\frac{\sqrt{2}}{2}}~~F~~D_{\rm PSF}~~
	({\frac{1}{57.3}}) ({\frac{1}{3600}})
\label{equ:psf_error}
\end{equation}
\noindent where $F$ is the focal length of the optic,
Table~\ref{tab:error_budget}.  This equation is useful
when specific models of the optic PSF are not available.

The above equation for $\sigma_{H}$ can be extended in two respects
given knowledge of the optic.  First there is generally a dependence
on energy which is slowly varying, thus $\sigma_{H}$ can be expressed
as, say, a polynomial in $\log_{10}(E)$.  Second, in the case of
\chandra, the PSF of the mirror shells is more cusp-like than
Gaussian. This cusp-like PSF causes the {\it effective} sigma of the
PSF projection to depend on the scale at which it is used, that is the
size of other error terms it is convolved with.
The following 
equations give approximations to the value of $\sigma_H$
for HETGS design purposes; blurs for the HEG and MEG
mirror sets are given separately:
\begin{equation}
  \sigma_{H,MEG} = 0.00998 ~+~ 0.00014\log_{10}E ~+~ -0.00399\log_{10}^2E
~+~ 0.000505\log_{10}^3E
\label{equ:hrma_sigma_meg}
\end{equation}
\begin{equation}
  \sigma_{H,HEG} = 0.01134 ~+~ 0.00675\log_{10}E ~+~ -0.01426\log_{10}^2E
~+~ 0.01133\log_{10}^3E
\label{equ:hrma_sigma_heg}
\end{equation}

\paragraph{Aspect Blur}
  Aspect reconstruction adds a blur that is expected to be of order $a$ =
0.34 arc seconds rms diameter for \chandra.  The resulting
one-dimensional rms sigma is thus:
\begin{equation}
  \sigma_{y'}, \sigma_{z'} = {\frac{1}{2}}
	{\frac{\sqrt{2}}{2}}~~F~~a~~({\frac{1}{57.3}})
({\frac{1}{3600}})
\label{equ:aspect_error}
\end{equation}
\noindent where $F$ is the HRMA focal length in mm.

\paragraph{Detector Pixel-size Blur}
  This error term is the spatial error introduced by the detector
readout scheme.  For a pixelated detector like ACIS we assume that the
PSF drifts with respect to the detector pixels and there is a uniform
distribution of the centroid location in pixel phase.  In this case
the reported location of an event is the center of the pixel when in
fact the event may have actually arrived $\pm 0.5$ pixel from the
center.  The rms value of such a uniform distribution is 0.29 times
the pixel size:
\begin{equation}
\sigma_{y'} = \sigma_{z'}= ~ 0.29 ~ L_{\rm pix}
\label{equ:pixel_error}
\end{equation}
\noindent If a uniform randomization of the pixel
value is applied during analysis, then a further uniform blur
is added in quadrature, adding a factor of $\sqrt{2}$.

\paragraph{Dither Rate Blur}
  A blur is added because the
arrival time of a photon at the ACIS detector is quantized
in units of a frame time.  The parameter $R_{\rm dither}$ is
the maximum dither rate expressed in units of arc seconds per
frame time and results in a blur term of:
\begin{equation}
  \sigma_{y'}, \sigma_{z'} = 0.29~~\frac{\sqrt{2}}{2}~~
	F~~R_{\rm dither}~~({\frac{1}{57.3}})({\frac{1}{3600}})
\label{equ:dither_error}
\end{equation}
\noindent where the factor of $\frac{\sqrt{2}}{2}$ is
present because the dither pattern is sinusoidal.

\paragraph{Defocus and Astigmatism Blurs}
  Including the effect of a defocus, $dx$, and a factor converting
the peak-to-peak blur into an rms equivalent, we get the following
equations for the Rowland astigmatism contribution to the error
budget in dispersion and cross-dispersion directions:
\begin{equation}
  \sigma_{y'} = 0.354~ {\frac{2R_0}{X_{\rm RS}}} ~dx
\label{equ:astig_dy}
\end{equation}
\begin{equation}
  \sigma_{z'} = 0.354~ {\frac{2R_0}{X_{\rm RS}}} ~(\Delta X_{\rm Rowland} + dx)
\label{equ:astig_dz}
\end{equation}
These equations assume that the detector conforms to the Rowland
circle except for an overall translation by $dx$ (positive towards the
HRMA).  The values of $R_0$ used in the error budget are effective
values -- weighted combinations of the relevant mirror shells.

\paragraph{Grating Period and Roll Variation Blurs}
  There are two main error terms which depend on how well the HETG
is built:  i) period variations within and between facets (``$dp/p$'')
and ii) alignment (``roll'') variations about the normal to the facet
surface within and between facets.  The period variations lead to
an additional blur in the dispersion direction:
\begin{equation}
  \sigma_{y'} \approx \beta ~ X_{\rm RS} ~ {\frac{dp}{p}}
\label{equ:dpop_error}
\end{equation}
\noindent where $dp/p$ is the rms period variation. 
The roll errors lead to additional blur
in the cross-dispersion direction through the equation:
\begin{equation}
  \sigma_{z'} \approx \beta ~ X_{\rm RS}
	~\gamma ~({\frac{1}{57.3}})(\frac{1}{60})
\label{equ:roll_error}
\end{equation}
\noindent where $\gamma$ is the rms roll variation in units of
arc minutes.



\clearpage



\clearpage

\clearpage

\clearpage

 \clearpage

 \clearpage
 
 \clearpage

 \clearpage 

\clearpage

\clearpage

\clearpage 

\begin{deluxetable}{lccl}
\tabletypesize{\scriptsize}
\tablecaption{Key fabrication, ground test, and flight parameters of
the HETG.\label{tbl-1}}
\tablewidth{0pt}
\tablehead{\colhead{Parameter Name} & \colhead{Value} & 
\colhead{Unit} & \colhead{Comments}
}
\startdata
~~~{\it Grating Facet Parameters:} & & & \\
Grating bar material & Gold & \nodata & \nodata \\
HEG, MEG Bar thickness & 510, 360 & nm & Approximate average value \\
HEG, MEG Bar width & 120, 208 & nm & '' \\
HEG, MEG polyimide thickness & 980, 550 & nm & ''\\
Plating base thicknesses & 20(Au), 5.0(Cr) & nm & ''\\
\tableline
~~~{\it HETG Laboratory Parameters:} & & & \\
HESS Rowland diameter & $8633.69 $ & mm & As designed and machined \\
HEG average period & $2000.81 \pm 0.05 $ & \AA & LR, NIST referenced \\
MEG average period & $[ 4001.41 \pm 0.22 ]$ & \AA & Updated in flight, see below. \\
Vignetting, shell 1 & $0.937 \pm 0.01$  & \nodata & Inter-facet vignetting, \\
Vignetting, shell 3 & $0.940 \pm 0.01$  & \nodata & '' from calculation.\\
Vignetting, shell 4 & $0.931 \pm 0.01$  & \nodata & '' \\
Vignetting, shell 6 & $0.936 \pm 0.01$  & \nodata & '' \\
Efficiencies (rev. N0004)& Figure~\ref{fig:hegmeg_effics} & 
  \nodata & from X-GEF measurements \\
 & & & and synchrotron optical constants \\
\tableline
~~~{\it XRCF Measurement Results:} & & & \\
Rowland Spacing at XRCF& $8782.8 \pm 0.6 $ & mm & Assuming lab periods \\
HEG angle & $-5.19 \pm0.05 $ & degree & w.r.t. XRCF axes \\
MEG angle &  $4.74 \pm 0.05 $ & degree &  '' \\
HEG--MEG opening angle & $9.934 \pm 0.008$ & degree & from beam center data \\
HEG $dp/p$ &  $146 \pm 50 $ &  ppm rms &  Mg-K slit scan analysis \\
MEG $dp/p$ &  $235 \pm 50 $ &  ppm rms &  '' \\
HEG roll variation &  $\approx 1.8$ & arc min. rms &  2 peaks, 3~arc~min. apart \\
MEG roll variation &  $\approx 1.8$ & arc min. rms &  $\approx$~Gaussian distribution\\
Mis-aligned MEGs  &  3~--~25 & arc min. &  6 MEG roll outliers \\
HEG contrib.~to LRF wing &  $\le 1.3\times 10^{-4}$ & 1/\AA~$\times$~\AA$^2$ 
     & at Mg-K, 9.887~\AA; see text. \\
MEG contrib.~to LRF wing &  $\le 2.0\times 10^{-4}$ & 1/\AA~$\times$~\AA$^2$ 
     &  '' \\
HEG scatter &  $\approx$~0.2 & \% /\AA & at 7~\AA ; $\le$~1~\% total \\
MEG scatter &  not seen & \nodata & $<$~1/10$^{\rm th}$ of HEG value \\
\tableline
~~~{\it Flight Results:} & & & \\
Rowland Spacing & $8632.65$ & mm & As-installed; sets wavelength scale \\
HEG angle & $-5.235 \pm0.01 $ & degree & w.r.t. ACIS-S3 CHIPX axis \\
MEG angle &  $4.725 \pm 0.01 $ & degree &  '' \\
HEG average period & $2000.81 \pm 0.05 $ & \AA & Retains ground cal.~value \\
MEG average period & $4001.95 \pm 0.13 $ & \AA & Based on Capella-HEG results \\
\tableline
\enddata




\end{deluxetable}

\clearpage

\begin{deluxetable}{lcrlcc}
\tabletypesize{\scriptsize}
\tablecaption{Simplified Resolving Power Error Budget.  The major
parameters and 
terms which contribute to the HETGS LRF blur are listed
here in a spreadsheet-like format.  The effective rms
contributions to the dispersion and cross-dispersion blur are given by
the referenced equations in Appendix~\ref{sec:rowland_append}.
As shown these blurs are rss'ed
together giving the size of the Gaussian LRF core
in each direction, $\sigma_{y'}^{\rm tot}$ and $\sigma_{z'}^{\rm tot}$.
The resolving power, $E/dE$, is also calculated as indicated.
Current flight parameter values are given here;
entries that differ for the MEG
and HEG gratings are shown as ``MEG\_value[HEG\_value]''.
\label{tab:error_budget}}
\tablewidth{0pt}
\tablehead{\colhead{Error Description} &
\colhead{Symbol} & 
\colhead{Value} & 
\colhead{Units} & 
\colhead{Dispersion Blur} & 
\colhead{Cross-Disp Blur} 
}
\startdata
~~~{\it Blur sources:} & & & & $\sigma_{y',i}$ & $\sigma_{z',i}$ \\
Optics PSF & $D_{\rm PSF}$ & $\approx$ 0.6 & arc sec rms dia.&
	Equ.~\ref{equ:psf_error} or
		\ref{equ:hrma_sigma_meg}[\ref{equ:hrma_sigma_heg}] &
	Equ.~\ref{equ:psf_error} or
		\ref{equ:hrma_sigma_meg}[\ref{equ:hrma_sigma_heg}] \\
Aspect & $a$ & $\approx$ 0.34 & arc sec rms dia.&
 	Equ.~\ref{equ:aspect_error} & Equ.~\ref{equ:aspect_error} \\
Detector pixel& $L_{\rm pix} $ & $0.023987$ & mm & 
	Equ.~\ref{equ:pixel_error} & Equ.~\ref{equ:pixel_error} \\
Dither rate & $R_{\rm dither}$ & 0.16 & arc sec/frame time &
	Equ.~\ref{equ:dither_error} & Equ.~\ref{equ:dither_error} \\
Defocus w/astig. & $dx$ & 0.1 & mm & 
	Equ.~\ref{equ:astig_dy} & Equ.~\ref{equ:astig_dz} \\
Period variation & $dp/p$ & 235[146] & $\times 10^{-6}$ rms &
	Equ.~\ref{equ:dpop_error} & ... \\
Roll variation & $\gamma$ & 1.8 & arc min. rms & 
	... & Equ.~\ref{equ:roll_error} \\
\tableline
 & & & {\it Total blur:} & $\sigma_{y'}^{\rm tot}=\sqrt{\sum_{i} \sigma_{y',i}^{2}}$ &
		$\sigma_{z'}^{\rm tot}=\sqrt{\sum_{i} \sigma_{z',i}^{2}}$ \\
 & & & {\it Resolving power:} &
	$ E/dE = {\beta X_{\rm RS}}/({{2.35\sigma_{y'}^{\rm tot}}}) $ & \\
 & & & & & \\
~~~{\it Input Parameters:} & & & & & \\
Energy & $E$ & as desired & keV & & \\
Period & $p$ & 4001.95[2000.81] & \AA & & \\
Effective Radius  & $R_0$ & 470.[330.] & mm & & \\
Rowland spacing  & $X_{\rm RS}$ & 8632.65 & mm & & \\
Focal length  &  $F$ & 10070.0 & mm & & \\
~~~{\it Derived Values:} & & & & & \\
Wavelength & $\lambda$ & 12.3985/E & \AA &  & \\
Diffr. angle & $\beta$ & $\arcsin (m\lambda /p)$ & radians &  & \\
Rowland offset & $\Delta X_{\rm Rowland}$ & $\beta^2 X_{\rm RS}$ & mm &  & \\
\tableline
\enddata

\end{deluxetable}
\clearpage

\end{document}